\documentclass[fleqn,usenatbib]{mnras}

\usepackage[T1]{fontenc}

\DeclareRobustCommand{\VAN}[3]{#2}
\let\VANthebibliography\thebibliography
\def\thebibliography{\DeclareRobustCommand{\VAN}[3]{##3}\VANthebibliography}

\usepackage{graphicx}	
\usepackage{amsmath}	
\usepackage{amssymb}	

\usepackage[normalem]{ulem} 

\newcommand{\bs}[1]{\boldsymbol{#1}}
\newcommand{\bcdot}{\boldsymbol{\cdot}}
\newcommand{\bnabla}{\boldsymbol{\nabla}}

\DeclareSymbolFont{matha}{OML}{txmi}{m}{it}
\DeclareMathSymbol{\varv}{\mathord}{matha}{29}

\usepackage{color}
\allowdisplaybreaks
\newcommand{\mat}{\ensuremath{\mathbfss}}

\title[Cosmic ray-driven galactic winds]{Cosmic ray-driven galactic winds: transport modes of cosmic rays and Alfv\'en-wave dark regions}

\author[T. Thomas et al.]{
T. Thomas,$^{1,2}$\thanks{E-mail: tthomas@aip.de (TT)}
C. Pfrommer,$^{1}$
R. Pakmor$^{3}$
\\
$^{1}$ Leibniz-Institute for Astrophysics Potsdam (AIP), An der Sternwarte 16, 14482 Potsdam, Germany\\
$^{2}$ Institute of Physics and Astronomy, University of Potsdam, Karl-Liebknecht-Str. 24/25, 14476 Golm, Germany\\
$^{3}$ Max Planck Institute for Astrophysics, Karl-Schwarzschild-Str. 1, 85741 Garching, Germany
}

\date{Accepted XXX. Received YYY; in original form ZZZ}

\pubyear{2020}

\begin{document}
\label{firstpage}
\pagerange{\pageref{firstpage}--\pageref{lastpage}}
\maketitle

\begin{abstract}
Feedback mediated by cosmic rays (CRs) is an important process in galaxy formation. 
Because CRs are long-lived and because they are transported along magnetic field lines independently of any gas flow, they can efficiently distribute their feedback energy within the galaxy. We present an in-depth investigation of (i) how CRs launch galactic winds from a disc that is forming in a $10^{11} \rmn{M}_\odot$ halo and (ii) how CR transport affects the dynamics in a galactic outflow. To this end, we use the \textsc{Arepo} moving-mesh code and model CR transport with the two-moment description of CR hydrodynamics. This model includes the CR interaction with gyroresonant Alfv\'en waves that enables us to self-consistently calculate the CR diffusion coefficient and CR transport speeds based on coarse-grained models for plasma physical effects. This delivers insight into key questions such as whether the effective CR transport is streaming-like or diffusive-like, how the CR diffusion coefficient and transport speed change inside the circumgalactic medium (CGM), and to what degree the two-moment approximation is needed to faithfully capture these effects. We find that the CR-diffusion coefficient reaches a steady-state in most environments with the notable exception of our newly discovered Alfv\'en-wave dark regions where the toroidal wind magnetic field is nearly perpendicular to the CR pressure gradient so that CRs are unable to excite gyroresonant Alfv\'en waves. However, CR transport itself cannot reach a steady-state and is not well described by either the CR streaming paradigm, the CR diffusion paradigm or a combination of both.
\end{abstract}



\section{Introduction}

CRs are charged particles that reach from non-relativistic to fully relativistic energies and span more than twelve orders of magnitude in energy. They constitute an important component inside the interstellar medium (ISM) because their energy density is comparable to that provided by turbulence, thermal pressure or magnetic pressure, respectively \citep{1990Boulares, 2005Cox, 2017Naab}. This already implies that CRs can impact the dynamics of the ISM in a significant way. CRs with energies $\gtrsim$~GeV are nearly collisionless and interact with their environment by scattering off of gyroresonant electro-magnetic waves \citep{BookSchlickeiser}, which couples CRs to the thermal plasma and enables them to exchange momentum and energy. 

Indeed, because of their ability to accelerate gas, CRs can launch, shape or alter galactic winds that influence the evolution of a galaxy \citep{2012Uhlig,2016PakmorII, 2017Ruszkowski, 2018Girichidis, 2018Farber, 2018Jacob, 2018Butsky, 2020Dashyan, 2022Quataert, 2022Farcy}. Equally important, CRs with approximately MeV energies that pervade the neutral gaseous phase of the ISM provide an important channel of ionisation contributing significantly to the molecular chemistry of the ISM \citep{2006Dalgarno, 2020Padovani}. After they are expelled into the CGM, CRs build up a pressure support against infalling gas whilst maintaining galactic outflows \citep{2016Salem, 2020Buck, 2020Li}. Models that account for CR acceleration within the jets of an active galactic nucleus (AGN) show a modified jet morphology and distribute the AGN feedback energy differently into the intracluster medium of the central regions of the cool core cluster \citep{2017Jacob,2019Yang,2018Ehlert}. High-energy CR ions are able to participate in hadronic interactions with the ambient gas, which produces e.g., gamma rays and radio-emitting electrons and positrons. Hence, improving our understanding of the spatial CR distribution is crucial for interpreting observations \citep{2008Evoli,2009Gabici, 2021WerhahnI, 2021WerhahnII, 2021WerhahnIII}.  

Owing to their quasi-collisionless nature, CRs show a distinct transport behaviour in every medium they are pervading. Magnetic fields are mainly mediating the dynamics and the transport of CRs. CR protons with GeV energies dominate the total energy budget of the CR population and interact with the magnetic field in the following way: (i) CRs propagate along the large-scale magnetic field because the CR gyroradius is small compared to typical astrophysical scales in galaxies and in the ISM  \citep{2013Zweibel,2017Zweibel,2021Hanasz}, (ii) CRs interact with magnetic fluctuations of the turbulent cascade through resonant and non-resonant interactions \citep{2004Yan,2011Yan,2022Lazarian}, and (iii) CRs themselves can drive small-scale magnetic perturbations by means of plasma physical processes and then interact with those magnetic fluctuations later on \citep{1969Kulsrud, 2021Shalaby, 2022Shalaby}.

Modelling CR dynamics is a difficult task because of the myriad of contributing physical processes, the large range of CR energies, and the large separation between length- and time-scales relevant for CRs and the astrophysical environment of interest. With the first advent of hydrodynamical theories for CR transport it was possible to show that CRs are able to drive galactic winds and can influence the structure of shocks \citep{1975Ipavich, 1986Drury, 1991Breitschwerdt, 2012Dorfi}. These hydrodynamical theories use a `one-moment' approximation for CR transport where the CR energy or number density is the only quantity that describes the entire CR population. These one-moment descriptions are frequently and successfully used in many applications. CR transport along magnetic fields in this approximation can be categorised using the CR diffusion or CR streaming pictures \citep{2013Wiener,2017Wiener}. If CRs are streaming, their interactions and momentum exchange with small-scale magnetic perturbations are sufficiently frequent so that the CRs are effectively comoving with the magnetic perturbations -- in the case of CRs scattering with Alfv\'en waves, the CRs population travels with the Alfv\'en speed. If scattering is less frequent, then CRs are not well coupled to magnetic perturbations and perform a random walk along magnetic field lines, which results in CR diffusion \citep{1971Skilling}. 

While the numerical modelling of CR diffusion is possible with only few modifications to standard numerical schemes, integrating CR streaming poses a challenge \citep{2009Sharma, 2011Sharma}. The requirement of a stable and accurate numerical scheme that is able to integrate the CR streaming equation resulted in the development of a second generation of hydrodynamical theories for CR transport that employ a two-moment approximation \citep{2018Jiang, 2019Thomas}. Therein, the momentum or flux density of CRs is additionally taken into account and provides a second moment of the CR population. This approach supersedes the previous one-moment description of CR transport as it does not any more rely on a predefined paradigm such as CR streaming or diffusion, but instead allows the emerging CR dynamics to freely adjust to changes, depending on the environmental properties. It is possible to improve the fidelity of spatial CR transport further by including higher-order moments but \citet{2022Thomas} showed that if scattering by magnetic perturbations is the dominant driver of CR transport, the two-moment description is sufficiently accurate in describing the resulting dynamics. 

In \citet{2020Thomas} the authors show that the combined CR streaming and diffusion transport model provided by a two-moment approximation, is in fact required to match the observed CR distribution of a class of non-thermal (radio harp) filaments located in the central molecular zone of our Galaxy that have been observed by the MeerKAT radio telescope \citep{2022Heywood}. A shortcoming of these two-moment hydrodynamical descriptions is their marginalisation over the entire CR energy spectrum so that they only provide a ``grey'' picture of CR hydrodynamics (CRHD). This neglects inherent differences between CR transport at different energies and different cooling processes acting on different parts of the CR energy spectrum. Transport descriptions that resolve the CR spectrum have been derived \citep{2020Girichidis, 2021Ogrodnik, 2022Hopkins} and show that e.g., galactic winds are launched differently in models employing either a grey or an energy-resolved description of CR transport \citep{2022Girichidis}.

In this work we use the grey two-moment method for CRHD of \citet{2019Thomas} that accounts for the CR interaction with self-generated Alfv\'en waves and evolves the energy densities of those Alfv\'en waves in addition to the energy and flux densities of CRs. The entire hydrodynamic subsystem of CR and Alfv\'en-wave equations is coupled to the thermal gas, which is described by the equations of magneto-hydrodynamics (MHD). We apply this model in the context of a CR-driven outflow of a disc galaxy, which forms inside a $M_{200} = 10^{11} \rmn{M}_\odot$ halo. We focus our discussion on how CRs and the galactic wind influence each other by examining (i) the exact wind launching mechanism that the CRs provide at the disc-halo interface and (ii) how CRs are transported in terms of their diffusion coefficients and realised streaming speeds. We demonstrate that neither CR diffusion nor CR streaming are appropriate descriptions for the resulting transport in our simulation. In addition, we study a new phenomenon called ``Alfv\'en wave dark regions'' within the galactic wind, which describes regions devoid of CR-generated Alfv\'en waves, and discuss the physical processes that lead to their formation. 

Our paper is structured as follows. In Section~\ref{sec:primer} we briefly recall those aspects of the CRHD model of \citet{2019Thomas} and the resulting CR transport dynamics that are relevant for this work. In Section~\ref{sec:setup} we detail our simulation setup and the used numerical parameters. We devote Section~\ref{sec:wind-launching} to a discussion of the properties and launching mechanism of a CR-driven galactic outflow. In Section~\ref{sec:energy_transport}, we investigate how CR energy is transported in the galactic outflow and study the emergent CR diffusion coefficient. Afterwards, we discuss the physical origin of Alfv\'en-wave dark regions in Section~\ref{sec:alfven-dark}.  We close this paper with a discussion of our results and possible caveats in Section~\ref{sec:discussion} and present our conclusion in Section~\ref{sec:conclusions}. We use the Heaviside-Lorentz system of units throughout this paper.

\section{Primer on CR Hydrodynamics}
\label{sec:primer}

In this section, we discuss aspects of CR hydrodynamics that will later on become important to understand the results of our simulation. 

\subsection{Reduced equations of CR hydrodynamics}

Instead of recalling the full set of CR hydrodynamical equations, here we focus our discussion on the transport of CRs along the magnetic field direction as well as the source terms due to gyroresonant CR interactions with Alfv\'en waves and their prevalent damping process, and refer to \citet{2019Thomas} for a discussion of all aspects of CR hydrodynamics. Their two-moment description for CR hydrodynamics follows the dynamics of CRs in terms of the energy density $\varepsilon_\mathrm{cr}$, its energy flux $f_\mathrm{cr}$ and the energy densities $\varepsilon_\mathrm{a,\pm}$ of gyroresonant Alfv\'en waves. These waves interact via gyroresonant CR scatterings as a result of the Lorentz force that is induced by plasma-scale magnetic perturbations of gyroresonant Alfv\'en waves. In \citet{2019Thomas} we self-consistently account for these scattering processes in terms of those four hydrodynamical quantities. In deriving the fundamental equations, we use the quasi-linear theory for the CR-Alfv\'en wave interactions and couple the CR fluid to the underlying thermal gas in an MHD approximation. In reduced form, which is sufficient for our discussion, these equations read:  
\begin{align}
    \frac{\partial \varepsilon_\mathrm{cr}}{\partial t} + \bnabla \bcdot (f_\mathrm{cr} \bs{b}) &= -\frac{\varv_\mathrm{a}}{3\kappa_+} [f_\mathrm{cr} - \varv_\mathrm{a}(\varepsilon_\mathrm{cr} + P_\mathrm{cr})] \nonumber \\ &\phantom{=}+\hspace{-6pt} \frac{\varv_\mathrm{a}}{3\kappa_-} [f_\mathrm{cr} + \varv_\mathrm{a}(\varepsilon_\mathrm{cr} + P_\mathrm{cr})] + \dots, \label{eq:energy_equation} \\
    \frac{\partial f_\mathrm{cr}}{\partial t} + c_\mathrm{red}^2\bs{b}\bcdot \bnabla  P_\mathrm{cr} &= -\frac{c_\mathrm{red}^2}{3\kappa_+} [f_\mathrm{cr} - \varv_\mathrm{a}(\varepsilon_\mathrm{cr} + P_\mathrm{cr})] \nonumber \\ &\phantom{=}\hspace{-6pt}+ \frac{c_\mathrm{red}^2}{3\kappa_-} [f_\mathrm{cr} + \varv_\mathrm{a}(\varepsilon_\mathrm{cr} + P_\mathrm{cr})]  + \dots, \label{eq:flux_equation} \\
    \frac{\partial \varepsilon_\mathrm{a,+}}{\partial t} + \bnabla \bcdot (\varepsilon_\mathrm{a,+} \varv_\mathrm{a} \bs{b}) &= +\frac{\varv_\mathrm{a}}{3\kappa_+} [f_\mathrm{cr} - \varv_\mathrm{a}(\varepsilon_\mathrm{cr} + P_\mathrm{cr})]\nonumber\\ &\phantom{=}- \alpha \varepsilon_\mathrm{a,+}^2 + \dots, \label{eq:forward_alfven_energy_equation}\\
    \frac{\partial \varepsilon_\mathrm{a,-}}{\partial t} + \bnabla \bcdot (\varepsilon_\mathrm{a,-} \varv_\mathrm{a} \bs{b}) &= -\frac{\varv_\mathrm{a}}{3\kappa_-} [f_\mathrm{cr} + \varv_\mathrm{a}(\varepsilon_\mathrm{cr} + P_\mathrm{cr})] \nonumber\\ &\phantom{=}- \alpha \varepsilon_\mathrm{a,-}^2 + \dots, \label{eq:backward_alfven_energy_equation}
\end{align}
where we intentionally omit the (advective and adiabatic) terms responsible for transport of CRs and gyroresonant Alfv\'en waves with the flow of the underlying gas, which is symbolised by the dots. 

The following quantities have been used: $\mathbfit{b} = \mathbfit{B} / B$ is the direction of the magnetic field $\mathbfit{B}$, which has a field strength $B$, $\varv_\mathrm{a}=B/\sqrt{\rho}$ is the Alfv\'en speed supported by the MHD fluid with a mass density $\rho$, and $c_\rmn{red}$ is the reduced speed of light, that is decreased from the physical value $c$ because of numerical efficiency reasons. However, we require $c_\rmn{red}$ to be much faster than the fastest MHD signal speed in the problem. The CR pressure is
\begin{align}
    P_\mathrm{cr} = (\gamma_\mathrm{cr} - 1) \varepsilon_\mathrm{cr} \quad \quad \text{where} \quad \quad \gamma_\mathrm{cr} = \frac{4}{3},
\end{align}
and $\varepsilon_\rmn{cr}$ denotes the CR energy density. The diffusion coefficients of CRs, $\kappa_\pm$, are defined by
\begin{align}
    \frac{1}{3\kappa_\pm} &= \frac{3\pi}{8} \frac{e B}{\gamma m c^3} \frac{\varepsilon_\mathrm{a,\pm}}{B^2}, \label{eq:def_kappa}
\end{align}
where $e$ is the elementary charge and $\gamma$ is the typical Lorentz factor of the CR protons of rest mass $m$. The diffusion coefficients are a measure for the coupling strength between CRs and gyroresonant Alfv\'en waves, and scale as $\kappa^{-1}_\pm \propto \nu_\pm\propto\varepsilon_\mathrm{a,\pm}$, where $\nu_\pm$ is the CR particle-wave scattering rate. We also include non-linear Landau damping of Alfv\'en waves, which enters Eqs.~\eqref{eq:forward_alfven_energy_equation} and \eqref{eq:backward_alfven_energy_equation} trough the terms $\alpha \varepsilon_\mathrm{a,\pm}^2$. The coupling constant for this process is given by
\begin{align}
\label{eq:eps_a,steady}
    \alpha &= \frac{\sqrt{\pi}}{8B^2} \frac{2eB}{\gamma m c^2} \sqrt{\frac{P_\mathrm{th}}{\rho}},
\end{align}
where $P_\mathrm{th}$ is the pressure of the thermal gas.

In \citet{2019Thomas}, we assume that the CRs are an ultra-relativistic fluid, which creates an interesting relationship between the CR energy flux density $f_\mathrm{cr}$ and the mean momentum density of the CR population: because ultra-relativistic particles have energies $E=\gamma m c^2$ and momenta $p\simeq \gamma m c$, we can interpret $f_\mathrm{cr} / c^2$ as the mean momentum density which simplifies the interpretation of our results.

\subsection{CR pressure and inertia}

We start by discussing the influence of the CR pressure term on the left-hand side of Eq.~\eqref{eq:flux_equation}. Due to the correspondence between CR energy flux and CR momentum density, this pressure term corresponds to a force. It is not a body force acting on the CR particles but a collective effect affecting a population of CRs. In the context of ideal thermal gases, the thermodynamic pressure is sustained by the molecular motion of individual gas particles in combination with collisions between gas particles. These particle-particle collisions are negligible for CRs. Here, the CR pressure term is purely sustained by the inertia of CRs. This inertia can collectively induce macroscopic fluxes of CRs in situations where CRs have an inhomogeneous distribution in space. We illustrate this with the following example: consider an isolated region of CRs without any initial CR flux. This situation can be realised by an equal amount of CRs travelling with and against the direction of the magnetic field. As CRs at both boundaries start to leave the region, this implies a non-zero flux of CRs outside the initial confinement region. In this scenario, there is no force acting on the CRs responsible for the resulting CR flux but simply the inertia of CRs in combination with an inhomogeneous CR distribution. 

In our simulations this CR inertia acts as a jump-start for the CR dynamics: there are more CRs inside a galactic disc in comparison to its galactic halo, which implies a CR pressure gradient that increases toward the disc. Hence, more CRs travel from the galactic disc towards the halo along a magnetic field line rather than the opposite way. This creates an initial flux of CRs pointing from the galactic disc to the halo. Once this flux exceeds the Alfv\'en speed, CRs start to excite gyroresonant Alfv\'en waves and their mutual interaction becomes important as we will discuss now.

\subsection{Gyroresonant interaction -- gas acceleration}

The gyroresonant interaction between CRs and small-scale Alfv\'en waves is described through the source terms on the right-hand sides of Eqs.~\eqref{eq:energy_equation} and \eqref{eq:flux_equation}. Microphysically, this process takes place if the gyromotion of CRs coincides with the gyration of the magnetic perturbation of small-scale Alfv\'en waves -- hence this is a resonant process. To discuss how energy is transferred by this process, we focus on CR interactions with  Alfv\'en waves that are propagating in the direction of the magnetic field. The first term on the right-hand side of Eq.~\eqref{eq:energy_equation} implies that if the velocity at which CR energy is transported,
\begin{equation}
    \varv_\mathrm{cr} = \frac{f_\mathrm{cr}}{\varepsilon_\mathrm{cr} + P_\mathrm{cr}}
\end{equation}
is larger in magnitude than the Alfv\'en speed, $|\varv_\mathrm{cr}|>\varv_\mathrm{a}$, and hence, is faster than Alfv\'en waves, CRs start to lose energy. As a result, the energy in forward travelling gyroresonant Alfv\'en waves increases at the same rate according to Eq.~\eqref{eq:forward_alfven_energy_equation}. The opposite process happens for $|\varv_\mathrm{cr}| < \varv_\mathrm{a}$, which implies CR energy gain by damping the energy contained in forward travelling gyroresonant Alfv\'en waves. The interaction between CRs and backward propagating Alfv\'en waves follows the same general idea. In summary, CRs lose energy to gyroresonant Alfv\'en waves by growing them once $|\varv_\mathrm{cr}| > \varv_\mathrm{a}$ and extract energy from them if $|\varv_\mathrm{cr}| < \varv_\mathrm{a}$. 

The CR flux density $f_\mathrm{cr}$ changes during this process so that it is increased (or decreased) towards $f_\mathrm{cr} = \pm \varv_\mathrm{a}(\varepsilon_\mathrm{cr} + P_\mathrm{cr})$ or $\varv_\mathrm{cr} = \pm \varv_\mathrm{a}$ as described by the source terms on the right-hand sides of Eq.~\eqref{eq:flux_equation}. This change corresponds to a mean force acting on the CR population because $f_\mathrm{cr}$ is an equivalent measure for the mean CR momenta. This force needs to be balanced by an opposing force to ensure momentum conservation. Because Alfv\'en waves are supported by the thermal plasma and because they are the source of gyroresonant interactions, they mediate this opposing force to the thermal gas. It is this process that enables CRs to accelerate the gas. In case of super-alfv\'enic CRs with $\varv_\mathrm{cr} > \varv_\mathrm{a}$, the interaction with forward travelling gyroresonant Alfv\'en waves effectively decelerates CRs (reduces $\varv_\mathrm{cr}$ or $f_\mathrm{cr}$), which in turn accelerates the thermal gas in the direction of the magnetic field.

\subsection{Damping of Alfv\'en waves -- gas heating}

CR-excited Alfv\'en waves can be damped by various processes such as collisionless non-linear Landau damping that we focus our attention on here. In this process two Alfv\'en waves with different wavenumbers non-linearly interact to create a beatwave, which in turn leads to small-scale pressure perturbations at the scale of the beatwave \citep{1991Miller}. This beatwave is subsequently Landau-damped and thus heats the gas. Thus, this process dissipates the energy contained in gyroresonant Alfv\'en waves by heating the underlying thermal gas. As CRs transfer energy to Alfv\'en waves and Alfv\'en waves dissipate their electromagnetic energy into heat, the combination of this damping process and the gyroresonant interaction opens up a channel for CRs to inject energy into their surrounding medium and to influence the thermodynamics therein.

\subsection{Steady state CR transport}

If the dynamical processes governing the transport of the CR fluid are sufficiently fast in comparison to any other relevant process, CR transport approaches a steady state which can be determined from Eq.~\eqref{eq:flux_equation} and which is characterised by the steady state flux:
\begin{align}
    f_\mathrm{cr, steady} &= \varv_\mathrm{st} (\varepsilon_\mathrm{cr} + P_\mathrm{cr}) - \kappa \bs{b} \bcdot \bnabla \varepsilon_\mathrm{cr}, \label{eq:steady_state_flux}
\end{align}
where
\begin{align}
    \kappa^{-1} &= \kappa^{-1}_+ + \kappa^{-1}_- \label{eq:total_diffusion_coefficient}
\end{align}
is the total diffusion coefficient and the CR streaming velocity is given by
\begin{align}
    \varv_\mathrm{st} &= \varv_\mathrm{a} \frac{\kappa^{-1}_+ - \kappa^{-1}_-}{\kappa^{-1}_+ + \kappa^{-1}_-}. \label{eq:steady_state_vst_kappa}
\end{align}
Whether it is possible to establish this steady state depends on the initial CR flux or velocity: the CRs must be faster than Alfv\'en waves, $|\varv_\mathrm{cr}| > \varv_\mathrm{a}$ because only in this case, the CRs themselves are able to continuously drive gyroresonant Alfv\'en waves that are required to maintain frequent particle-wave scatterings and to eventually approach this steady state. In the other case, $|\varv_\mathrm{cr}| < \varv_\mathrm{a}$, the CRs damp gyroresonant Alfv\'en waves and diminish residual Alfv\'en wave energy, which reduces the scattering rate, and consequently increases the dynamical timescale of CRs. Because the Alfv\'en wave energy is drained exponentially fast, it is not possible to reach the steady state without an external source of gyroresonant Alfv\'en waves. 

Provided CR energy is transported faster than Alfv\'en waves, $|\varv_\mathrm{cr}| > \varv_\mathrm{a}$, only one of the two types of gyroresonant Alfv\'en waves can be excited: because CRs are faster than one type of Alfv\'en waves, they are automatically slower than the other wave type. In this case $\varv_\mathrm{st} = \pm \varv_\mathrm{a}$ and the velocity difference between CRs and Alfv\'en waves in Eq.~\eqref{eq:steady_state_flux} is given by the diffusion term. The sign of this velocity difference determines which type of gyroresonant Alfv\'en wave is excited: e.g., if we assume $\bs{b} \bcdot \bnabla P_\mathrm{cr} < 0$, then CR inertia accelerates the CR population so that they are faster than those Alfv\'en waves that propagate along the direction of the magnetic field, i.e., $\varv_\mathrm{cr} > \varv_\mathrm{a}$ and enables those waves to grow, implying $\varepsilon_\mathrm{a,+} \neq 0$ and $\kappa_+ < \infty$. For $\bs{b} \bcdot \bnabla P_\mathrm{cr} > 0$, we obtain $\varepsilon_\mathrm{a,-} \neq 0$ and $\kappa_- < \infty$. If one of the wave types is being damped, only one diffusion coefficient is not infinite, and the expression in Eq.~\eqref{eq:steady_state_vst_kappa} simplifies to the negative sign of the CR energy gradient or
\begin{align}
    \varv_\mathrm{st,steady} &= - \varv_\mathrm{a} \mathrm{sign}\left(\bs{b} \bcdot \bnabla \varepsilon_\mathrm{cr}\right).
\end{align}
Thus, in steady state CRs preferentially ``stream down their gradient''.

\section{Simulation Setup}
\label{sec:setup}

We perform a CRHD simulation of an isolated disc galaxy with the moving-mesh code \textsc{Arepo} \citep{2010Springel}. The galaxy is initialized in a $M_{200} = 10^{11} \rmn{M}_\odot$ halo with $10^6$ equal-mass gas particles particles that are in hydrostatic equilibrium with the dark matter halo. The setup up is similar to those used in previous works \citep{2016PakmorII,2017bPfrommer,2021Pfrommer,2021WerhahnI}. The dark matter halo is static and follows the analytic form of an NFW halo with concentration parameter $c = 7$ \citep{1997Navarro}. Using dark matter particles to simulate a dynamically evolving dark matter halo has little impact on the overall galaxy evolution in this setup \citep{2018Jacob}. Gas cells are initialised following the same spherically symmetric NFW profile with a baryon mass fraction of $\Omega_\rmn{b} / \Omega_\rmn{m} =0.155$. Halo masses and length scales are measured in units of $h^{-1} \rmn{M}_\odot$ and $h^{-1} \rmn{kpc}$, where $H_0 = 100\,h~\mathrm{km}~\mathrm{s}^{-1}\,\mathrm{Mpc}^{-1}$is today's value of the Hubble constant. Initially, the gaseous halo slowly rotates with a total angular momentum of $ J = \lambda \, G M_{200}^{5/2} / \sqrt{E_{200}}$ where $G$ is Newton's constant, $E_{200}$ is the energy contained in the system, defined by equation~(22) in \citet{1998Mo}, and $\lambda = 0.05$ is the dimensionless spin-parameter. The initial magnetic field is pointing along the $x$ axis with a uniform magnetic field strength of $10^{-10}$G. Neither CRs nor gyroresonant Alfv\'en waves are present in the initial conditions.

The equations of two-moment CRHD are discretised with the finite volume scheme as described in \citet{2021Thomas}, which is an extension of the existing and well-tested CR and MHD modules of \textsc{Arepo} \citep{2013Pakmor, 2017Pfrommer} and which uses the second-order time integrator and reconstruction algorithms of \citet{2016Pakmor}. The CRHD equations of \citet{2021Thomas} use the P1 Eddington approximation as the closure relation that was shown to provide robust and converged results in comparison to higher order closure schemes \citep{2022Thomas}. Optically thin gas cooling, star formation, and the adopted effective equation of state used to model the ISM are described in \citet{2003Springel}.

CRs are accelerated at the shocks of supernova (SN) remnants and escape later into the ISM. In order to model this process, we inject CR energy
\begin{equation}
    E_\mathrm{CR, inj} = \zeta e_\mathrm{SN} m_\star
\end{equation}
into the neighbouring $32 \pm 1$ computational cells weighted with a spline kernel (as employed in smoothed particle hydrodynamics) instantaneously after a star particle with mass $m_\star$ was created. The parameter $\zeta = 0.05$ defines the fraction of kinetic SN energy that is converted into CR energy. The specific energy $e_\mathrm{SN} = 10^{49} \mathrm{erg}\, \mathrm{M}_\odot^{-1}$ is the approximate average energy that is released by core-collapse SNe per solar mass of newly formed stars assuming a Kroupa initial mass function and $10^{51} \mathrm{erg}$ per supernova \citep{2001Kroupa, 2017Pfrommer}. Note that we do not directly inject thermal or kinetic SN energy into the ISM because this process is already modeled by the effective equation of state. In addition to CRs, SN shocks also amplify the magnetic field in their surroundings. Instabilities driven by CRs in the precursor of these shocks such as the Bell or smaller-scale instabilities \citep{2004Bell,2021Shalaby,2022Shalaby} create small-scale turbulent magnetic fields that are able to scatter CRs. The creation of these magnetic fields can be described by a local gain of $\varepsilon_\mathrm{a,\pm}$ around SN remnants. Because (i) the fraction of energy contained in the Alfv\'enic mode of MHD turbulence in those precursors and (ii) the fraction of Alfv\'en waves that are able to effectively escape the SNe site are two unknowns, we inject a conservative amount of energy in gyroresonant Alfv\'en waves, 
\begin{align}
    E_\mathrm{a, +, inj} &= 10^{-5} E_\mathrm{CR, inj},
    \quad\mbox{and}\\
    E_\mathrm{a, -, inj} &= 10^{-5} E_\mathrm{CR, inj}
\end{align}
together and in the same fashion as CRs for every newly created star particle.

\textsc{Arepo} provides the ability to refine and derefine computational cells based on arbitrary user-defined criteria. While the standard Lagrangian refinement scheme keeps the mass enclosed by a computational cell nearly constant, super-Lagrangian refinement schemes can be used to increase the resolution of the simulation in regions of interest \citep{2019Suresh,2019Hummels,2019vandeVoort}. Here, we focus our attention on the dynamics and the launching of the galactic outflow and thus apply a super-Lagrangian refinement to gas cells situated in the (inner) CGM. We define the following mass and volume targets for computational cells:
\begin{itemize}
    \item Cells should have a target mass $M_\mathrm{cell, target} = 1.55 \times 10^{4} \mathrm{M}_\odot$,
    \item Cells at galactocentric radii $r < 60$~kpc should have a target volume $V_\mathrm{cell, target} = 4 \pi r_\mathrm{cell, target}^3 / 3$ with a maximum equivalent radius $r_\mathrm{cell, target} = 600$pc,
    \item Cells at galactocentric radii $r < 30$~kpc should have a maximum equivalent radius $r_\mathrm{cell, target} = 200$pc,
    \item Cells at galactocentric radii $r < 2.5$~kpc should have a maximum equivalent radius $r_\mathrm{cell, target} = 50$pc.
\end{itemize}
If either cell mass or volume do not match their target values within a factor of two, the cell is either refined or de-refined in order to again fulfil the criteria. Furthermore, to avoid large volume deviations of adjacent cells we refine a computational cell if one of its Voronoi neighbours has a $10$ times smaller volume.

For the CR transport we use a reduced speed of light of $c_\mathrm{red} = 8000~\mathrm{km}\,\mathrm{s}^{-1}$ and subcycle the CR transport along the magnetic field lines 8 times for each invocation of the MHD solver. Because typical sound and Alfv\'en speeds reach at maximum of $1000~\mathrm{km}\,\mathrm{s}^{-1}$, our choice of the reduced speed of light leads to well separated light-like and MHD characteristics. CR cooling is implemented such that  $f_\mathrm{cr} / \varepsilon_\mathrm{cr}$ and thus the effective CR transport velocity is kept constant during CR cooling. We use the CR Coulomb and hadronic cooling rates provided by \citet{2017Pfrommer}.

\section{CR-driven Outflows}
\label{sec:wind-launching}

In this section, we first assess the general properties of CR driven outflows using our two-moment method of CRHD and then focus on the exact driving mechanism of the galactic wind in our simulation.

\subsection{Outflow properties}

The initial evolution of the simulation follows those of other simulations of isolated disc galaxies: the gas loses pressure support by radiative cooling and starts to fall towards the centre of the halo. While conserving the total angular momentum, a gaseous disc starts to form inside out. As the first gas cells start to exceed the star formation threshold,  their gas mass is converted into star particles, which inject their feedback energy into their surroundings. Initially, there is no outflow for the first Gyr and the disc is only able to launch an bipolar galactic wind once the injected CR energy provides enough pressure in the central regions. This behaviour is similar to other simulations that use a collapsing gaseous halo setup, which are discussed in detail in previous works \citep{2016PakmorII,2017bPfrommer,2018Jacob}. In this section we discuss general properties of this outflow and the hydrodynamical processes that lead to its launching.

\begin{figure*}
	\includegraphics[width=\textwidth]{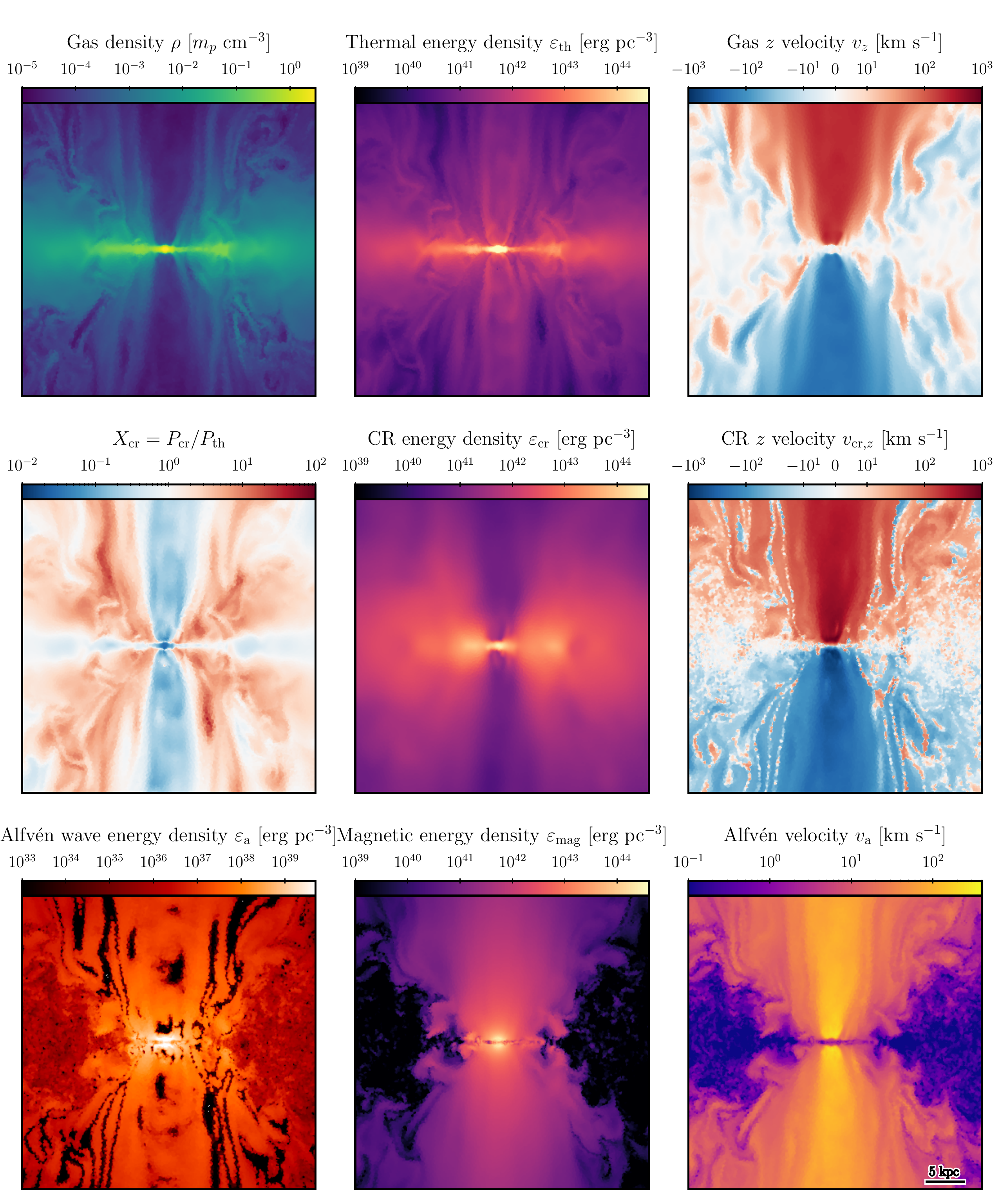}
	\caption{Gallery showing various quantities of the galactic outflow in quadratic slices with $40$~kpc side length laying in the $x$--$z$ plane centred on the centre of the simulation box at $t=4$~Gyr. The panels for the thermal, CR, and magnetic energy densities share the same dynamic range while the panel for the Alfv\'en wave energy density uses a different scale.}
    \label{fig:gallery}
\end{figure*}

The emerging outflow is bipolar, has gas velocities $\sim 100$ km s$^{-1}$ and is launched in the central kpc of the galactic disc as can be inferred from Fig.~\ref{fig:gallery} at $t=4$~Gyr. The morphology of this outflow resembles the structure of AGN jets as it consist of an inner fast and underdense outflow (the spine in the AGN jet terminology) and an outer, slower-moving cylindrical shell that shears the stationary CGM at larger radii (the sheath in AGN jet terminology). 

The inner outflow is generally characterised by a high Alfv\'en velocity that reaches $\varv_\mathrm{a} \sim 1000$ km s$^{-1}$ near the galactic plane. This high Alfv\'en velocity is due to the interplay of two processes: (i) the wind ejects gas from the galactic plane and carries along magnetic field lines which were amplified by the magnetic dynamo to high magnetic field strengths and (ii) the inner outflow has a low density. Both properties are required to explain the high Alfv\'en velocity in the inner outflow (bottom-right panel of Fig.~\ref{fig:gallery}) because the magnetic energy density and thus in the magnetic field strength does not show the spine-sheath structure and instead shows a smoothly declining strength in the CGM towards larger cylindrical radii (bottom-middle panel). 

The inner wind also shows a depression in CR energy density (central panel of Fig.~\ref{fig:gallery}) and a slight enhancement of thermal energy density (top middle panel). The central outflow is dominated by thermal energy, which can be inferred from the map of $X_\mathrm{cr} = P_\mathrm{cr} / P_\mathrm{th}$ (centre right panel). This quantity prominently changes from $X_\mathrm{cr} > 1$ (CR pressure dominated) in the outer CGM to $X_\mathrm{cr} < 1$ in the inner outflow. CRs are generally transported away from the disc with speeds comparable to the Alfv\'en velocity, which is visible in the projected $z$ component of the CR velocity defined by $\varv_{\mathrm{cr}, z} = \varv_\mathrm{cr} \mathbfit{b} \cdot \mathbfit{e}_z$, where $\mathbfit{e}_z$ is the unit vector pointing in $z$ direction.

The outer wind is CR pressure dominated and has a more structured morphology in gas density, thermal and Alfv\'en wave energy densities, as well as in all three velocity variables shown in Fig.~\ref{fig:gallery} (Alfv\'en velocity, $z$ components of the gas and CR velocities, right column). The outermost boundary of the outflow is traced by the sharp transition in the CGM material that is enriched by magnetic fields with energy densities $\varepsilon_\mathrm{mag} \gtrsim 10 ^{40}$~erg pc$^{-3}$ and CGM material that only shows spurious enhancements of the magnetic field strength. This transition region is also characterised by a patchy structure in the vertical component of the gas velocity: some of the gas is currently falling towards the disc while in other places, gas is moving away from the mid plane. Gas in the outer wind can shear in this transition region via Kelvin-Helmholtz type instabilities and mix with the rest of the CGM. As a result, this creates the more diverse morphology of the outer wind in comparison to the seemingly laminar inner outflow.

The bottom-left panel of Fig.~\ref{fig:gallery} shows the total energy density contained in Alfv\'en waves, $\varepsilon_\mathrm{a} = \varepsilon_\mathrm{a,+} + \varepsilon_\mathrm{a,-}$. This quantity also follows the morphology of the wind: $\varepsilon_\mathrm{a}$ has high values (10$^{39}$ erg pc$^{-3}$) immediately above and below the galactic disc, strongly decreases in the galactic outflow (10$^{37}$ erg pc$^{-3}$) and attains its lowest values (10$^{36}$ erg pc$^{-3}$) in the rest of the CGM. Furthermore, there are distinctive filamentary structures where $\varepsilon_\mathrm{a} \sim 0$ in isolated regions, so called ``Alfv\'en-wave dark regions''. While these regions are patchy within the inner wind, they are vertically elongated in the outer wind. These regions in the inner wind do not correlate with any other displayed quantity. By contrast, the elongated dark regions in the outer wind can also be found at places where the $z$-component of the CR velocity $\varv_{\mathrm{cr}, z}$ changes sign. Because the wind is otherwise filled with Alfv\'en waves, the most likely reason for the absence of Alfv\'en wave in these regions are localised damping processes: the sign-reversal of $\varv_{\mathrm{cr}, z}$ points towards low CR velocities themselves, indicating more slowly moving CRs in comparison to the fast Alfv\'en waves, which would then be effectively damped through the Fermi-II process. We further discuss the origin of these ``Alfv\'en-wave dark regions'' in Section~\ref{sec:alfven-dark}.

\begin{figure*}
	\includegraphics[width=0.975\linewidth]{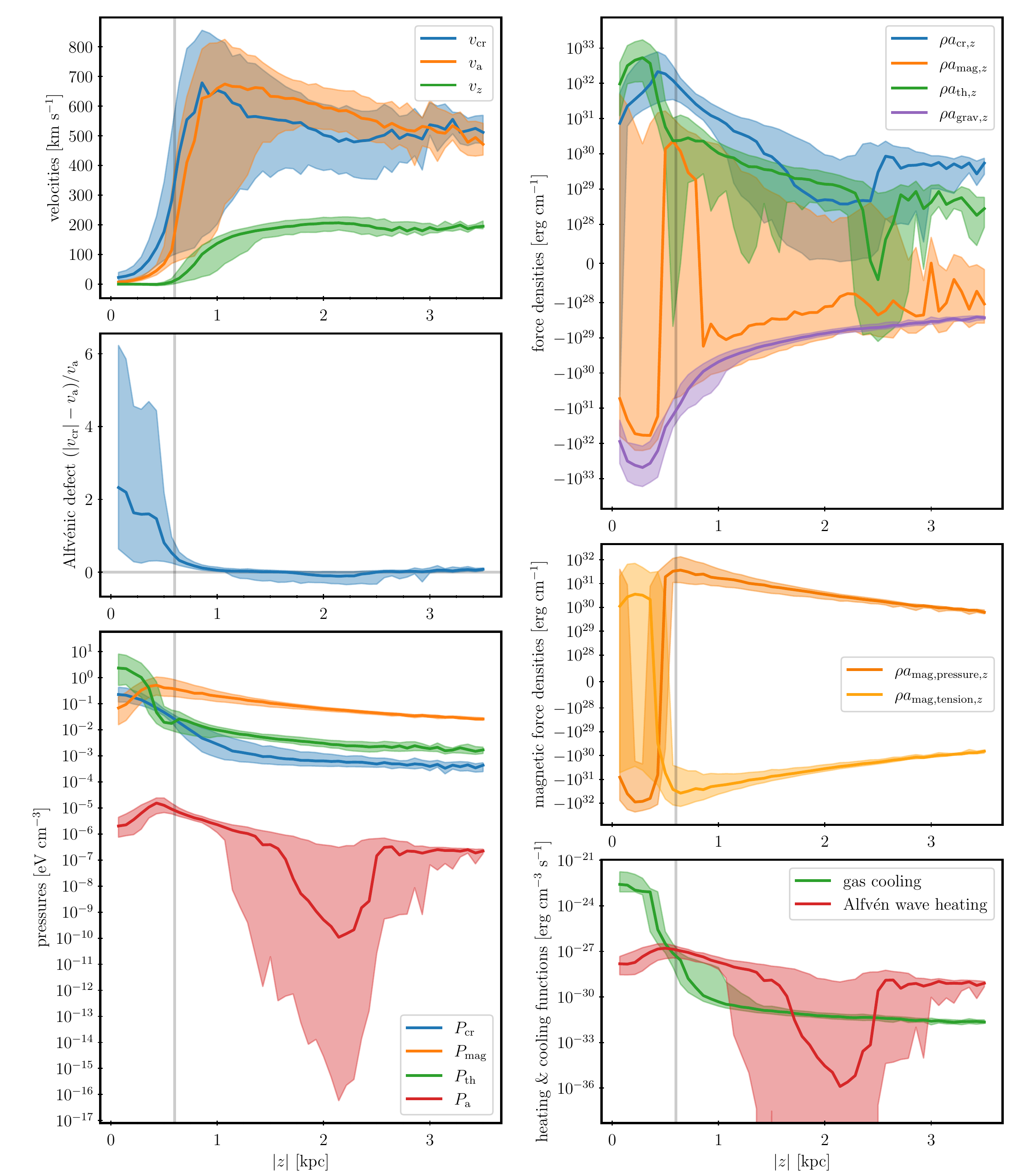}
	\caption{Quantities relevant for the discussion of the wind-launching mechanism measured in a cylindrical region defined by $|z| < 3.5$~kpc and $R < 1$~kpc at $t = 4$~Gyr. We sample this region with 50 equally spaced vertical bins and display the median (coloured lines) and the 20$^{\rmn{th}}$-to-80$^{\rmn{th}}$ percentiles (shaded regions) of each volume-weighted quantity inside each bin. The top-left panel shows the Alfv\'en and CR velocity (for energy transport), as well as the $z$ component of the gas velocity. The middle left panel shows the dimensionless Alfv\'en defect $(|\varv_\mathrm{cr}| - \varv_\mathrm{a}) / \varv_\mathrm{a}$, which is discussed in the main-text. The lower left panel shows the CR, thermal, magnetic and Alf\'en wave pressures while the upper right panel shows the force densities exerted on the gas by the magnetic field, CRs, gravity, and the thermal pressure. The force densities of the magnetic field are separated into contributions from the magnetic tension and pressure forces in the middle-right panel. The lower right panel shows the gas cooling and heating rates originating from optically thin cooling processes and Alfv\'en wave heating. The grey vertical line marks $|z|=0.6$~kpc, the wind launching site.}
    \label{fig:inner_profiles}
\end{figure*}

\subsection{Wind launching}

The outflow is launched by CRs at the disc-halo interface. One of our super-Lagrangian refinement requirements imposes an upper limit on the effective cell radii of $50$~pc to properly resolve this region in our simulation. The aim of such a high-resolution region is to ensure a large number of computational cells that provide a sufficiently good statistics to accurately investigate the wind launching mechanism. In Fig.~\ref{fig:inner_profiles}, we display relevant quantities in slices of cylindrical region defined by $|z| < 3.5$~kpc and $R = \sqrt{x^2 + y^2} < 1$~kpc which encloses the wind launching site. In the top-left panel, we display the Alfv\'en and CR velocity (for energy transport), as well as the $z$ component of the gas velocity. Within the galactic disc, for $|z| < 0.5$~kpc, CRs move faster than the local Alfv\'en speed. Both start to share similar velocities once the wind is launched. This happens at $|z| \sim 0.6$~kpc, the height at which the gas is accelerated to move away from the galactic disc. This acceleration continues up to a galactic height $|z|\sim 1$~kpc where $|\varv_{z}|\sim 150$ km s$^{-1}$. At larger galactic heights the acceleration becomes weaker until the maximum outflow velocity, $|\varv_{z}|\sim 200$ km s$^{-1}$ is reached at $|z|\sim 2$~kpc. 

\begin{figure}
	\includegraphics[width=\linewidth]{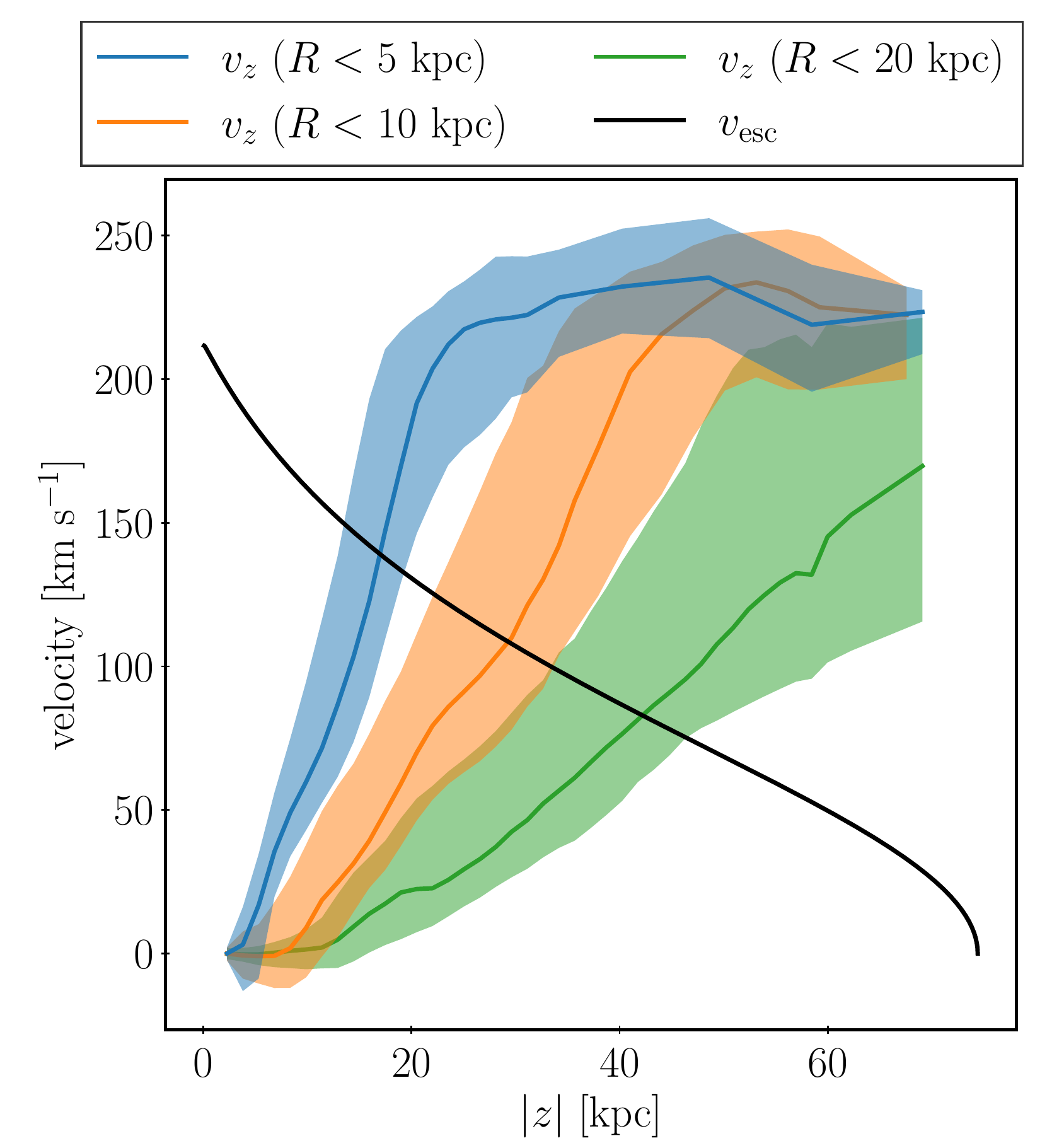}
	\caption{Median (coloured lines) and 20$^{\rmn{th}}$-to-80$^{\rmn{th}}$ mass-weighted percentiles (shaded regions) of the vertical velocity are compared to the escape velocity of the initial NFW halo as a function of vertical height at $t = 4$~Gyr. We sample $\varv_z$ with 50 equally spaced bins from $z = 0$~kpc to  $|z| = 70$~kpc and only consider cells with $R < 5$, 10, and 20~kpc, respectively. To increase our statistics, we merge neighbouring vertical bins until each bin contains a minimum of 1000 computational cells. Note that we flip the sign of $\varv_z$ for gas particles with $z<0$ so that the profiles shows outflow velocities with respect to the disc.}
	\label{fig:wind_velocity}
\end{figure}

Using the Alfv\'enic transport defect $(|\varv_\mathrm{cr}| - \varv_\mathrm{a}) / \varv_\mathrm{a}$ (middle-left panel of Fig.~\ref{fig:inner_profiles}) we can understand how this outflow is launched. This quantity is equal to zero if the CR energy is transported with the local Alfv\'en velocity and larger if the CR energy is transported faster than the Alfv\'en velocity. As discussed in Section~\ref{sec:primer}, CRs have the potential to accelerate the ambient gas if this Alfv\'enic defect is larger than zero. If the Alfv\'enic defect is zero, CRs and Alfv\'en waves are tightly coupled and CRs are exerting a pressure force on the ambient gas. Below the wind launching side, CR energy is transported with approximately two times the Alfv\'en velocity which decreases for increasing $z$ indicating that CRs and Alfv\'en waves start to couple. Both velocities start to converge just at the wind launching site, leaving CRs and the gas tightly coupled. Thus, at the wind launching site CRs are able to accelerate the ambient gas. Note that also the statistical scatter (displayed by the 20th to 80th volume-weighted percentiles) of the Alfv\'enic defect is low around the wind launching side which points towards a coherent process at this location. 

The ability of CRs to accelerate the gas does not imply that CRs are actually responsible for the wind launching. Magnetic or thermal pressures could also lift the gas into an outflow. To discuss this possibility we show the relevant pressures in the bottom-left panel of Fig.~\ref{fig:inner_profiles}. A steep pressure gradient in $P_\mathrm{th}$ or $P_\mathrm{mag}$ at the wind launching site would indicate that the wind could also be pressure driven. The thermal pressure profile shows two turning points in the vicinity of the wind launching site whereas both, the CR and magnetic pressure profiles have gradients that can lead to an effective vertical acceleration. To quantify the overall contribution of the forces caused by magnetic fields, CRs, thermal pressure, and gravity to the final acceleration of the gas, we display the vertical components of
\begin{align}
    \rho \mathbfit{a}_{\rmn{mag}} &= \rho \mathbfit{a}_{\rmn{mag},\rmn{pressure}} + \rho \mathbfit{a}_{\rmn{mag},\rmn{tension}} \\
    &=- \bnabla \bcdot \left(\frac{B^2}{2} \mat{1} - \mathbfit{B}\mathbfit{B}\right), \\
    \rho \mathbfit{a}_{\rmn{cr}} &= - \bnabla_\perp P_\rmn{cr} + \frac{\mathbfit{b}}{3\kappa_+} [f_\rmn{cr} - \varv_\rmn{a} (\varepsilon_\rmn{cr} + P_\rmn{cr})]  \\
    &\phantom{=} + \frac{\mathbfit{b}}{3\kappa_-} [f_\rmn{cr} + \varv_\rmn{a} (\varepsilon_\rmn{cr} + P_\rmn{cr})],\\
    \rho \mathbfit{a}_{\rmn{th}} &= - \bnabla P_\rmn{th},
\end{align}
in the top-right panel of Fig.~\ref{fig:inner_profiles} together with the vertical force density caused by gravity, $\rho \mathbfit{a}_{\rmn{grav}}$ (that includes the external gravitational potential and self-gravity of the gas). CRs clearly dominate the outward-pointing force near the wind launching site. The force exerted by the thermal pressure comes second while it is nearly two orders of magnitude weaker than the force attributed to CRs. The total force exerted by the magnetic field shows alternating acceleration directions. It accelerates the gas away from the discs only at the wind launching side. Otherwise the median force points towards the galactic disc and counteracts the outflow. This may seem counter-intuitive because the magnetic pressure gradient in the top-left panel of Fig.~\ref{fig:inner_profiles} shows that the magnetic pressure force is accelerating gas in the outflow direction at and above the wind launching side. To resolve this puzzle, we separately show the contribution from magnetic pressure and tension forces in the middle-right panel of Fig.~\ref{fig:inner_profiles}. The magnetic pressure force profile follows our expectations from the median pressure profile: gas is accelerated away from disc right at and above the wind launching side. But the magnetic tension force accelerates the gas in the opposite direction in this region. Because the magnetic tension and pressure forces have comparable magnitudes but different signs, their combined net force shows a smaller magnitude. This cancellation effect is not perfect and the sign and magnitude of the residual net magnetic force varies spatially, which explains the alternating signs of the net magnetic force densities.

We display the heating and cooling rates that govern the thermodynamics of the gas in the wind in the bottom-right panel of Fig.~\ref{fig:inner_profiles}. Gas heating is provided through damping of gyroresonant Alfv\'en waves. Consequently, this heating is strong at the wind launching site where CRs efficiently excite Alfv\'en waves through the streaming instability. Because Alfv\'en-wave heating exceeds gas cooling above the wind-launching site, the gas is effectively heated. This explains the increase of thermal pressure in this region and causes the generation of turning points (see bottom-left panel of Fig.~\ref{fig:inner_profiles}). At heights $1~\rmn{kpc}\lesssim |z| \lesssim 2.5$~kpc, CRs propagate slower than Alfv\'en waves (see top- and middle-left panel) and the streaming instability ceases to transfer energy from CRs to Alfv\'en waves, which suffer unbalanced strong (non-linear Landau) damping. This leads to low values of the energy contained in Alfv\'en waves and the associated pressure in this region. Because of the decreasing level of Alfv\'en waves, there is less energy available for damping, which implies a lower heating rate in this region. Furthermore, we see a reduction of the CR mediated force because the availability of Alfv\'en waves influences the coupling strength between CRs and the gas. Gas with the lowest median Alfv\'en-wave heating rate at $|z|\sim 2$~kpc shows a reduced thermal pressure force possibly because in that region, the median gas cooling rate locally exceeds the Alfv\'en wave heating rate.

In Fig.~\ref{fig:wind_velocity}, we display the median and the 20$^{\rmn{th}}$-to-80$^{\rmn{th}}$ mass-weighted percentiles of the vertical gas velocity profile, which we compare to the escape velocity $\varv_\mathrm{esc}$ defined by
\begin{equation}
    \varv_\mathrm{esc}^2(z) = 2 \int_z^{R_\mathrm{200}} \mathrm{d}\,r \frac{G M_{<r}}{r^2},
\end{equation}
where $M_{<r}$ is the mass contained in a sphere of radius $r$ within the initial NFW halo and $R_{200} = 74.4~ \rmn{kpc}$ is the virial radius. To compute these velocity profiles, we only consider gas cells at galactocentric (cylindrical) radii of $R < 5$, 10, and 20~kpc. We observe a continuous increase of the outflow velocity for all three cylinder radii. Note that both, the entrained halo gas of the wind and the gas that is part of the gaseous halo count toward the mass-weighted statistics. Consequently, a larger velocity value at a larger elevations does not necessary imply that all of the CGM at this height has been accelerated but instead means that the mass of the CGM that is not part of the outflow loses statistical weight. Furthermore, for larger cylindrical radii $R$, there is more halo gas included in the statistics, which causes shallower gradients in the outflow velocity profiles. Nevertheless, the outflow velocity exceeds the escape velocity of the halo in all cases, indicating that CR feedback is in principle able to expel the ISM beyond the virial radius and to substantially reduce the amount of gas in the galaxy that is necessary for ongoing star formation.

\section{CR Energy Transport}
\label{sec:energy_transport}
In this section, we discuss how the speed of CR energy transport, $\varv_\mathrm{cr}$, compares to other characteristic velocities and how its simulated value compares to values derived from the steady-state assumption. Second, we assess how well the CR diffusion coefficient, which is a measure of the CR scattering rate and directly related to the energy density of gyroresonant Alfv\'en waves, compares to the steady-state approximation. We emphasise that the transport velocity of the CR energy and the diffusion coefficient are different and complementary aspects of CR transport, which are not trivially related in the two-moment picture of CRHD. By contrast, the CR transport velocity respectively its flux and the CR diffusion coefficient are directly related in the case of the one-moment description of CRHD via Eq.~\eqref{eq:steady_state_flux}.

\subsection{CR transport velocity}
\label{sec:transport_velocity}

\begin{figure}
	\includegraphics[width=\linewidth]{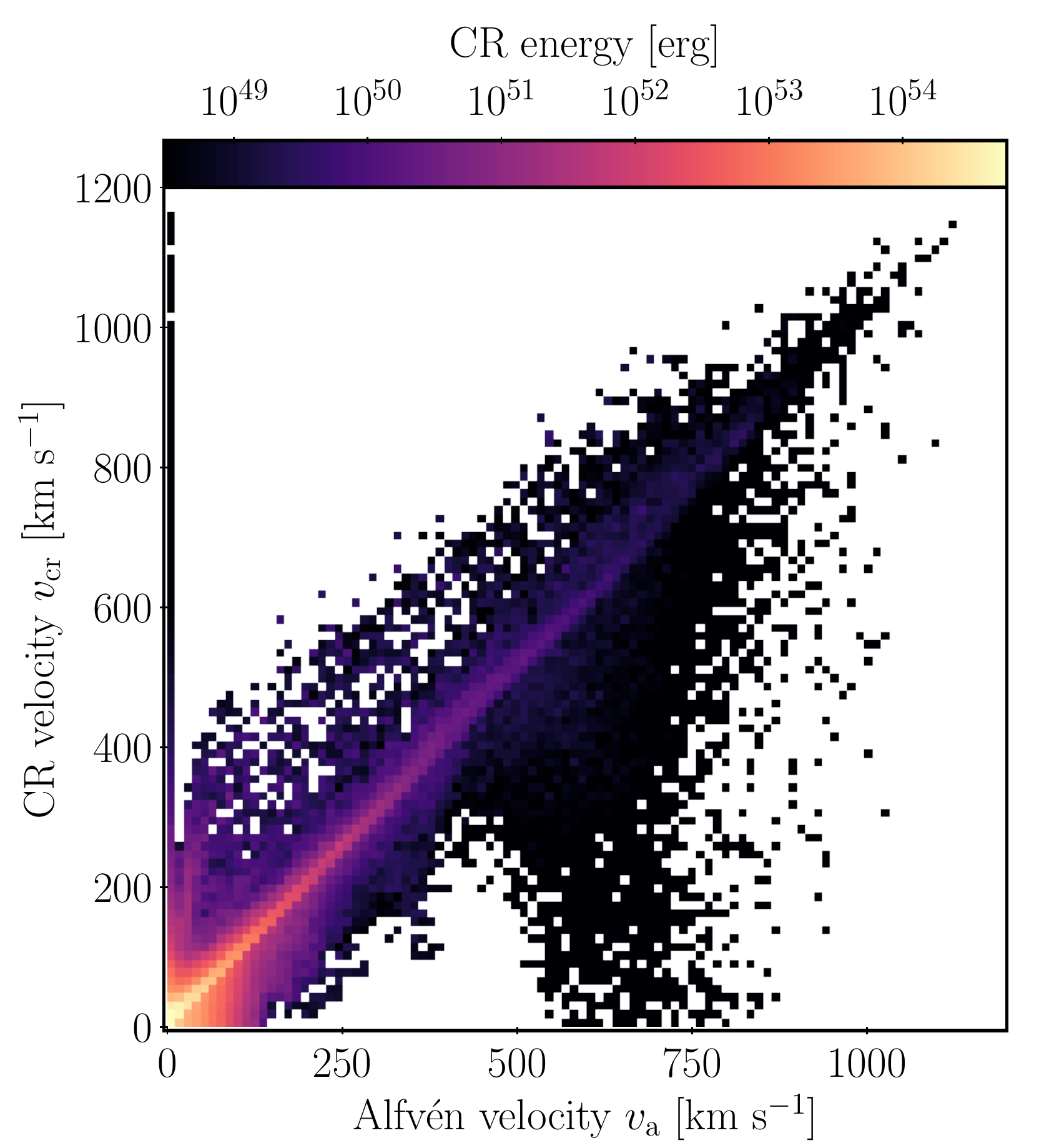}
	\caption{Two-dimensional histogram of the plane spanned by CR velocity and Alfv\'en speed at $t = 4$~Gyr where we weight each computational cell of the simulation with its CR energy. The correlation between both quantities shows that CR energy is mostly transported with the local Alfv\'en velocity.}
	\label{fig:streaming}
\end{figure}

In the one-moment approximation for CR transport the evolution of the CR energy is described by a convection-diffusion process, where the convection of CRs energy is realised by CRs streaming down their gradient. The characteristic speed for this process is given by the Alfv\'en speed. In Fig.~\ref{fig:streaming}, we display the $|\varv_\mathrm{cr}|$--$\varv_\mathrm{a}$ distribution for the whole simulation box with a CR-energy weighting. The most striking feature in this distribution is the prominent accumulation of energy on 1-to-1 relation. This shows that most of the CR energy is transported at velocities comparable to the Alfv\'en velocity. Deviation from the 1-to-1 relation are either created by the influence of CR diffusion or can be explained by a non steady-state transport of CRs. CR diffusion, the second cornerstone of the convection-diffusion picture, can increase $\varv_\mathrm{cr}$ and thus explain the fraction of CR energy above the 1-to-1 relation. One can notice that the 1-to-1 relation is indeed extended towards higher CR transport velocities in the range $\varv_\mathrm{a} \sim 300$--$600$~km s$^{-1}$ while it has a clear cutoff towards lower values of $\varv_\mathrm{cr}$. Thus CR diffusion can explain the deviations in this range. 

CR energy that is transported slower than Alfv\'en waves, $\varv_\mathrm{cr} \lesssim \varv_\mathrm{a}$, cannot be explained in the convection-diffusion picture. A large amount of CR energy situated at $\varv_\mathrm{a} < 200$~km s$^{-1}$ falls into this category and is mainly derived from the galactic disc where these low Alfv\'en velocities can be found in combination with a considerable amount of CR energy. The last noticeable feature is the extended region at $\varv_\mathrm{a} \sim 500$--$800$~km s$^{-1}$ where CR energy is transported slower than Alfv\'en waves. In this region, the Fermi-II process is able to operate and thus, we expect the corresponding spatial regions to have a small amount of Alfv\'en wave energy. Small values of $\varv_\mathrm{cr}/\varv_\mathrm{a}<1$ and  high Alfv\'en speeds can be found within the Alfv\'en wave dark regions in the galactic outflow (visible in the bottom-left panel of Fig.~\ref{fig:gallery} as dark patches and filaments).

\begin{figure}
	\includegraphics[width=\linewidth]{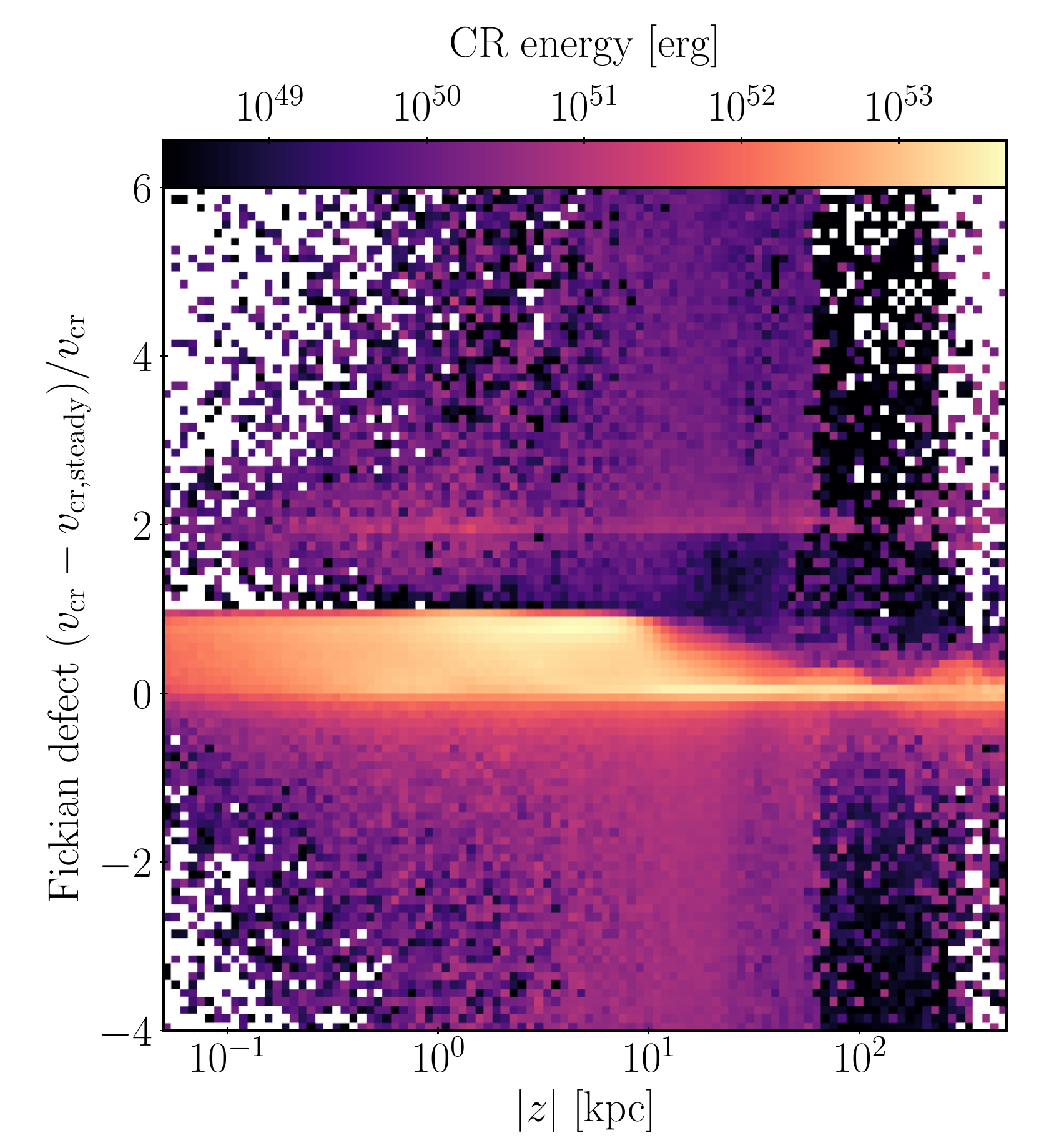}
	\caption{Two dimensional histogram of the Fickian defect $D_\mathrm{F} = (\varv_\mathrm{cr} - \varv_\mathrm{cr, steady}) / \varv_\mathrm{cr}$,  which represents the dimensionless deviation of the CR velocity to its predicted steady state value $\varv_\mathrm{cr, steady}$ in a streaming-diffusion approximation, at $t = 4$~Gyr. A zero value of $D_\mathrm{F}$ means that the CR energy is transported with its steady state velocity; a value of one means that CR energy is transported although the steady state assumption predicts no transport; a value of two means that CR energy is transported with the steady state velocity but opposite to its predicted direction. We weight each computational cell of the simulation with its CR energy before binning into this histogram.}
	\label{fig:fickian_defect}
\end{figure}

We further quantify the deviation of the simulated CR transport and the steady-state transport by comparing the steady-state velocity, defined by
\begin{equation}
    \varv_\mathrm{cr, steady} = \frac{f_\mathrm{cr, steady}}{\varepsilon_\mathrm{cr} + P_\mathrm{cr}} = \varv_\mathrm{st} - \kappa \frac{\bs{b} \bcdot \bnabla \varepsilon_\mathrm{cr}}{\varepsilon_\mathrm{cr} + P_\mathrm{cr}}, \label{eq:vcr_steady}
\end{equation}
with the simulated CR energy transport velocity using the Fickian defect
\begin{equation}
    D_\mathrm{F} = \frac{\varv_\mathrm{cr} - \varv_\mathrm{cr, steady}}{\varv_\mathrm{cr}}.
\end{equation}
We plot the CR-energy weighted distribution of this quantity with respect to the galactic height $z$ in Fig.~\ref{fig:fickian_defect}. We use the CR diffusion coefficient as calculated by Eq.~\eqref{eq:total_diffusion_coefficient} and hence based upon the local energy density of Alfv\'en waves. The Fickian defect can be interpreted as follows:
\begin{itemize}
    \item If $D_\mathrm{F} = 0$ then the CR energy transport velocity coincides with its value from the steady-state approximation.
    \item If $D_\mathrm{F} = 1$ then the CR energy transport velocity in steady-state approximation is negligible compared to the realized velocity. Both values differ considerably.
    \item If $D_\mathrm{F} = 2$ then both the CR energy transport velocity in steady-state approximation and its simulated counterpart coincide in magnitude but have opposite signs. Both values differ considerably.
    \item If $D_\mathrm{F} < 0$ then the CR energy is transported slower than predicted by the steady-state approximation.
\end{itemize}
Because most of the CR energy is located in the galactic disc for galactic heights $|z| < 1$, we find no clear correlation of the simulated and steady-state CR transport velocity in the ISM. Most CRs in this region have a Fickian defect of $0 \lesssim D_\mathrm{F} \lesssim 1$ so that CR energy is transported with its steady-state velocity or faster. Because of these substantial deviations it is thus hard to argue in favour of the steady-state approximation, which appears not to be applicable in the ISM. The same holds true in the inner CGM defined here by $1 < |z| < 10$, where there is an even smaller fraction of CRs transported at the steady-state velocity. The galactic wind expels CRs into parts of the outer CGM with galactic heights $|z| > 10$. These regions dominate the statistics in the corresponding part of Fig.~\ref{fig:fickian_defect}. There we mainly find $D_\mathrm{F} \approx 0$ implying that CRs transport their energy at the steady-state velocity which is a good approximation in the outflow region of the outer CGM.

\begin{figure}
    \centering
	\includegraphics[width=0.8\linewidth]{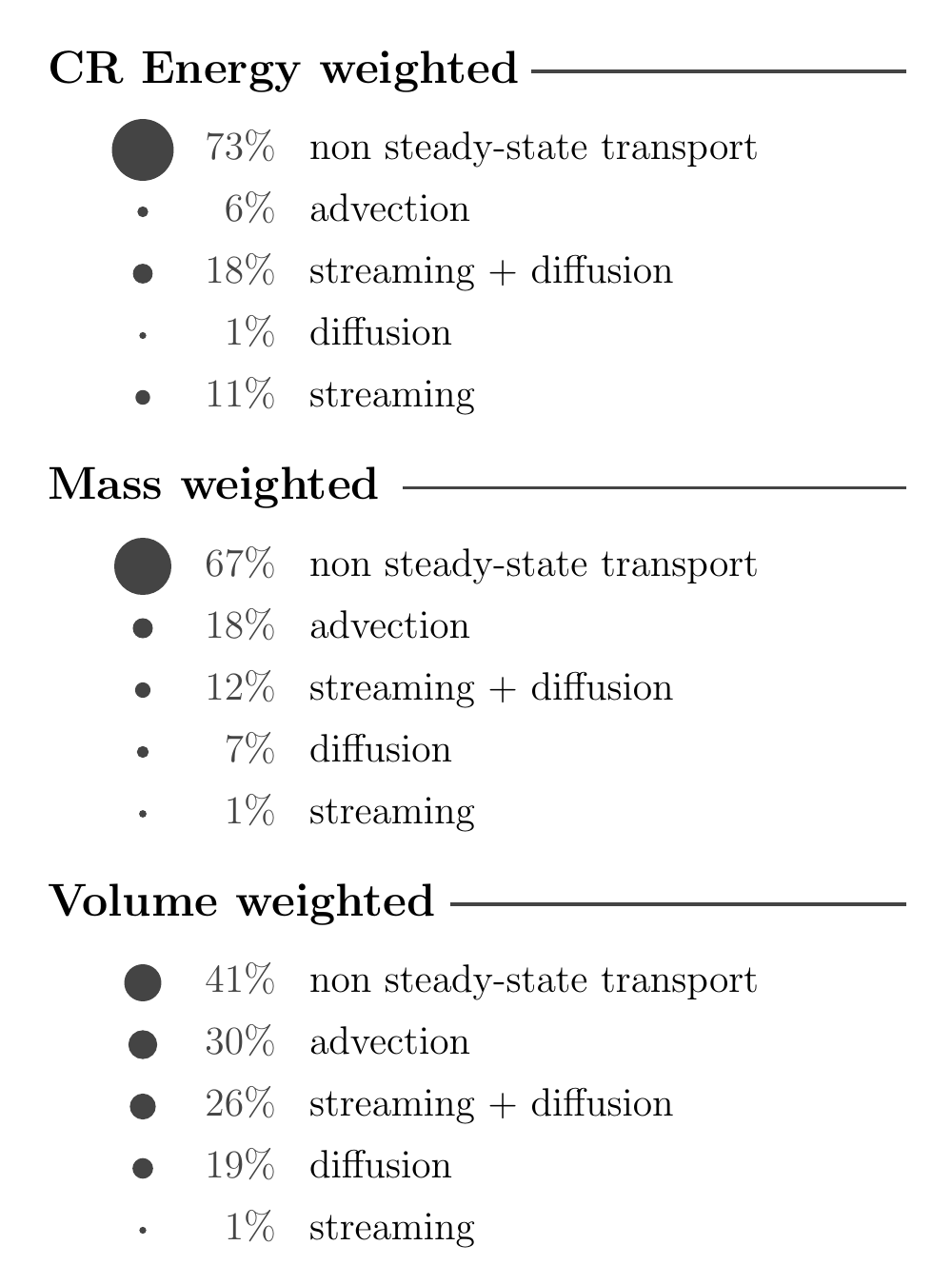}
	\caption{Classification of the realised CR transport mode into categories representing CR transport paradigms with different statistical weighting schemes. The categories are defined in the main text. Note that by our definitions the categories 'advection', 'streaming + diffusion', 'diffusion', and 'streaming' share some overlap in case the CR transport is slow. If a computational cell does not fulfil at least one of the criteria be to characterised by one of the steady-state transport modes then it counts towards the 'non steady-state transport' category.}
	\label{fig:velocity_classification}
\end{figure}

Given that the streaming-diffusion picture is only valid in some parts of the simulation domain, we are going to quantify the CR transport into various categories. To this end, we compare the CR energy transport velocity as realised in our simulation to the steady-state prediction. We define the following categories using conditions that we apply to the computational cells in the simulation: 
\begin{itemize}
    \item a computational cell can be described by the pure streaming model if $|\varv_\mathrm{cr} - \varv_\mathrm{st}| < 0.1 |\varv_\mathrm{cr}|$, which defines the `streaming' category, 
    \item a computational cell can be described by the pure diffusion model if $|\varv_\mathrm{cr} - \kappa (\bs{b} \bcdot \bnabla \varepsilon_\mathrm{cr}) / (\varepsilon_\mathrm{cr} + P_\mathrm{cr})| < 0.1 |\varv_\mathrm{cr}|$, which defines the `diffusion' category,
    \item a computational cell can be described by the mixed streaming-diffusion model if $|\varv_\mathrm{cr} - \varv_\mathrm{cr, steady}| < 0.1 |\varv_\mathrm{cr}|$, which defines the `streaming + diffusion' category,
    \item a computational cell can be described by the advection model if $|\varv_\mathrm{cr}| < 0.1 |\varv_\mathrm{cr, steady}|$, which defines the `advection' category, and
    \item if a computational cell does not match the criteria of any of the previous categories and thus CR transport cannot be described by a steady-state model, it is labelled by the `non steady-state transport' category.
\end{itemize}

This categorisation requires some clarifications and discussions on possible overlapping categories. First, the `streaming + diffusion' category contains but is not limited to all pure `streaming' and pure `diffusion' cases so that it is equal or larger than the sum of the two individual categories. Second, a given computational cell can fall into multiple categories describing steady-state transport:  (i) if the local Alfv\'en velocity or the velocity associated with CR diffusion is small, then the transport in these cells can also be described by CR advection, and (ii) if the Alfv\'en and CR diffusion velocities are comparable by coincidence, then this computational cell contributes to both the streaming and diffusion categories. Third, the 'non steady-state transport' has, by definition, no overlap with any of the other categories. In conclusion, the categories represent the consistency of CR transport in our simulation with one or the other transport paradigm.

Figure~\ref{fig:velocity_classification} displays the result of this categorisation for different statistical weightings. The weighting is accomplished by summing up the statistical weights of the individual cells that contribute to a category and divide afterwards by the total weight of the whole simulation to get the quoted percentages. As noted above, because the categories are not mutually exclusive, the total sum of percentages in one weighting scheme necessarily exceeds the 100 per cent mark. For all weightings explored in Fig.~\ref{fig:velocity_classification}, the non steady-state transport is the most populated category and thus, most of the time CR transport is inconsistent with any steady-state approximation. This highlights the importance of the two-moment approximation, which enables the description of CR transport beyond the steady-state paradigm. If we were to abandon the two-moment approximation and would like to describe CR transport with a steady-state approach,  then the most important CR transport processes to be considered would be the 'advection' and 'streaming + diffusion' categories.

\subsection{CR diffusion coefficient}
\label{sec:diffusion-coefficient}

Although we have just shown that the pure diffusion approximation is inconsistent with most of CR transport in our simulation, the concept of a CR diffusion coefficient is nevertheless relevant for this discussion: it is not only a measure for the transport speed in the diffusion approximation but is particularly a measure of the CRs scattering rate $\nu_\pm \propto \kappa_\pm^{-1}$ with gyroresonant Alfv\'en waves and as such, determines how tightly CRs are coupled to the thermal gas via Alfv\'en-wave scattering.

\begin{figure}
	\includegraphics[width=\linewidth]{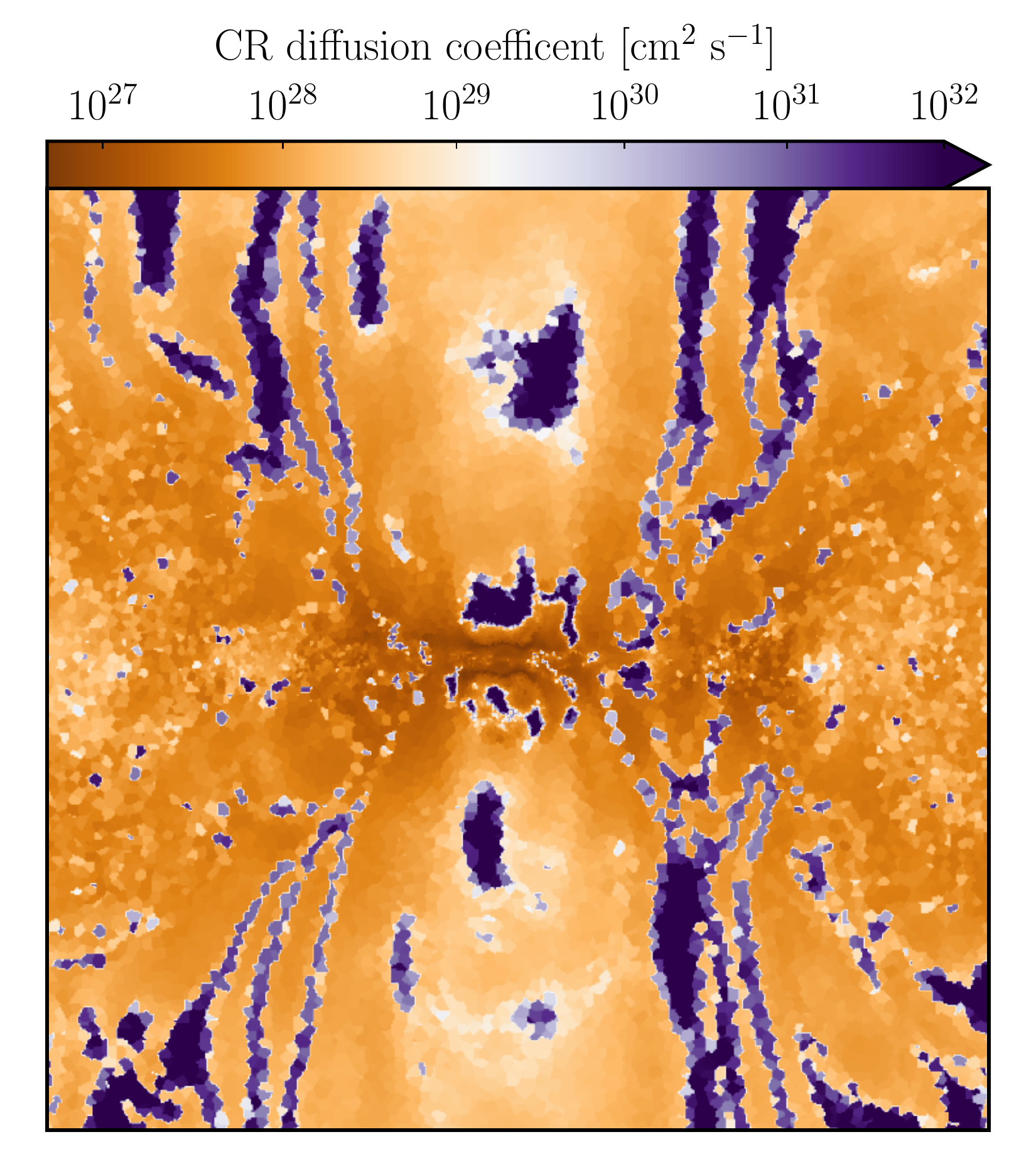}
	\caption{(Total) CR diffusion coefficient $\kappa$ (as defined in Eq.~\ref{eq:total_diffusion_coefficient}) in a slice through the simulation domain at $t = 4$~Gyr. The figure extent and slice positioning are as in Fig.~\ref{fig:gallery}.}
	\label{fig:kappa_gallery}
\end{figure}
Figure~\ref{fig:kappa_gallery} shows the total diffusion coefficient in the same slice through the simulation box as used in Fig.~\ref{fig:gallery}. Overall the diffusion coefficient ranges from $\kappa\sim 3 \times 10^{27} \mathrm{cm^2} \mathrm{s}^{-1}$ to $\sim 10^{29} \mathrm{cm^2} \mathrm{s}^{-1}$ in the inner CGM with the notable exceptions of the Alfv\'en wave dark regions in the outflow (where $\kappa\to\infty$).  
The values of $\kappa$ show a general gradient that is aligned with the direction of the outflow. We find lower values of the diffusion coefficient within the galactic disc and larger values in the outflow. The low diffusion coefficient in the galactic disc is caused by the absence of strong damping -- the non-linear Landau damping mechanism is efficient where the energy density of Alfv\'en waves and temperatures are high. While there are sufficient gyroresonant Alfv\'en waves in the disc, the ISM in our simulation is warm, which lowers the overall effectiveness of this damping mechanism. Note that ion-neutral damping, which we neglected in the present work, also plays an important role in setting the diffusion coefficient in the ISM. 

In the outflow, non-linear Landau damping is active due to high temperatures and the presence of Alfv\'en waves. Yet, we attribute the gradient of the diffusion coefficient not entirely to the damping of Alfv\'en waves but also to (i) the prevalent adiabatic losses that are stronger in comparison to the ISM and (ii) to the slower amplification of Alfv\'en waves at larger galactic heights. The increased rate of adiabatic losses are a direct consequence of the stratification of $\varepsilon_\mathrm{cr}$ and $B$ in the outflow (see Fig.~\ref{fig:gallery} and \ref{fig:inner_profiles}). The lower magnetic and CR energy densities imply a smaller driving of gyroresonant Alfv\'en waves because the self-confinement terms on the right-hand sides of Eqs.~\eqref{eq:forward_alfven_energy_equation} and \eqref{eq:backward_alfven_energy_equation} scale as $\propto B \varepsilon_\mathrm{cr} (|\varv_\mathrm{cr}| - \varv_\mathrm{a})$. CRs have large diffusion coefficients in these Alfv\'en wave dark regions because $1 / \kappa \propto (\epsilon_{\mathrm{a}, +} + \epsilon_{\mathrm{a}, -})$, which states that the absence of gyroresonant Alfv\'en waves implies a large diffusion coefficient. 

Because we only account for Alfv\'en waves that are created through the streaming instability and neglect other dynamical sources such as Alfv\'en waves that result from turbulent cascading, we underestimate the possibly available energy in form of magnetic fluctuations in Alfv\'en wave dark regions. Hence the derived values of the CR diffusion coefficient may overestimate the true value that results from CR-scattering with all forms of magnetic fluctuations. Consequently, the presented diffusion coefficient in Alfv\'en wave dark regions should be regarded as an upper limit.

Previous studies routinely assume that the CR diffusion coefficient can be approximated by a steady-state approximation where the growth and damping of gyroresonant Alfv\'en waves balance each other. We can check the validity of this approximation using our simulation. To calculate the steady state, we 
\begin{figure}
	\includegraphics[width=\linewidth]{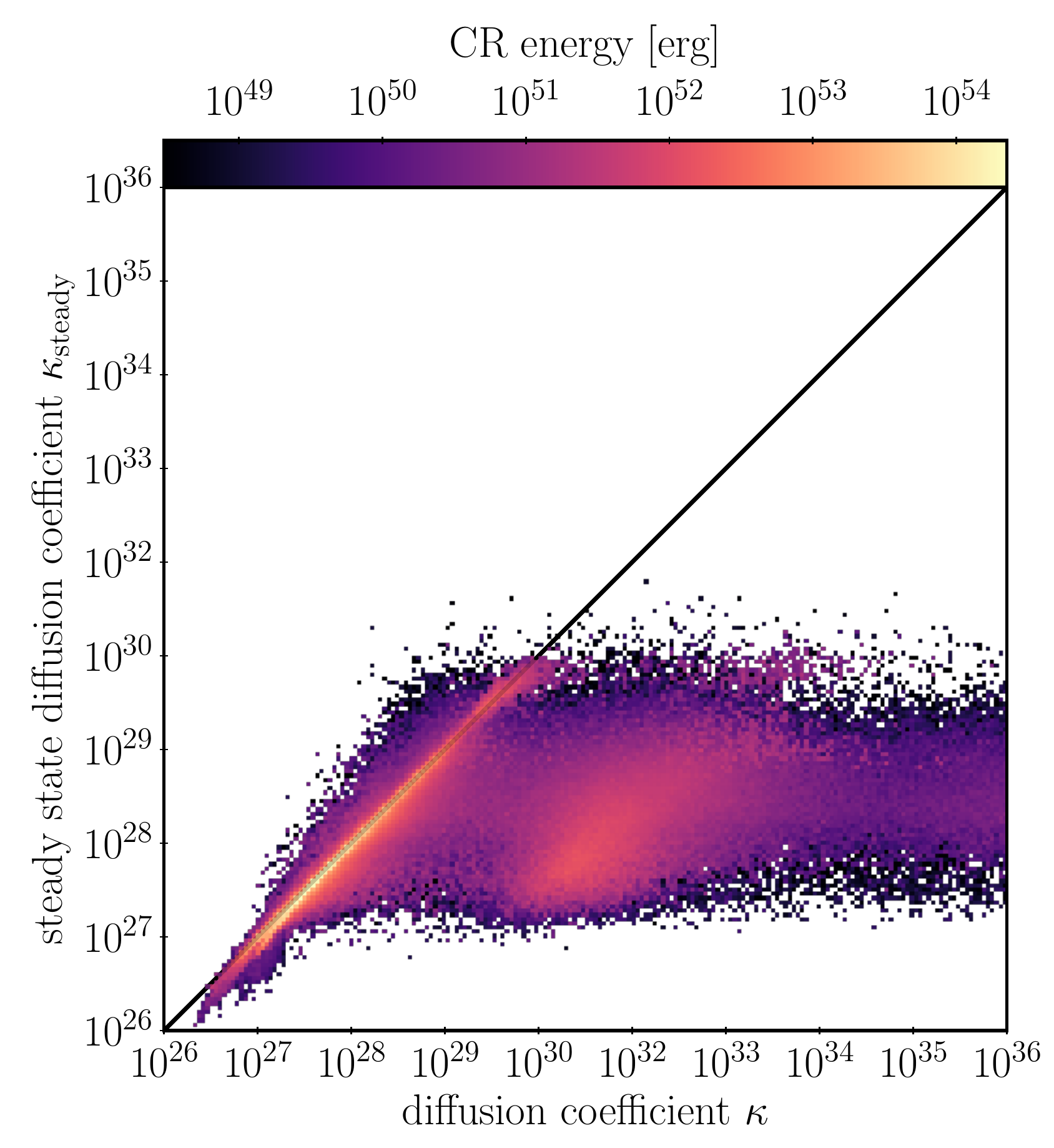}
	\caption{Comparison of the simulated diffusion coefficient (horizontal axis, defined in Eq.~\ref{eq:total_diffusion_coefficient}) and the steady state diffusion coefficient (vertical axis, defined in Eq.~\ref{eq:kappa_steady}) at $t = 4$~Gyr. The steady state diffusion coefficient is calculated based on the assumed instantaneous balance between Alfv\'en wave growth and damping. We weight each computational cell of the simulation with its CR energy before binning into this histogram. The grey line corresponds to a 1-to-1 relation. }
	\label{fig:steady_kappa}
\end{figure}   
\begin{figure*}
	\includegraphics[width=\linewidth]{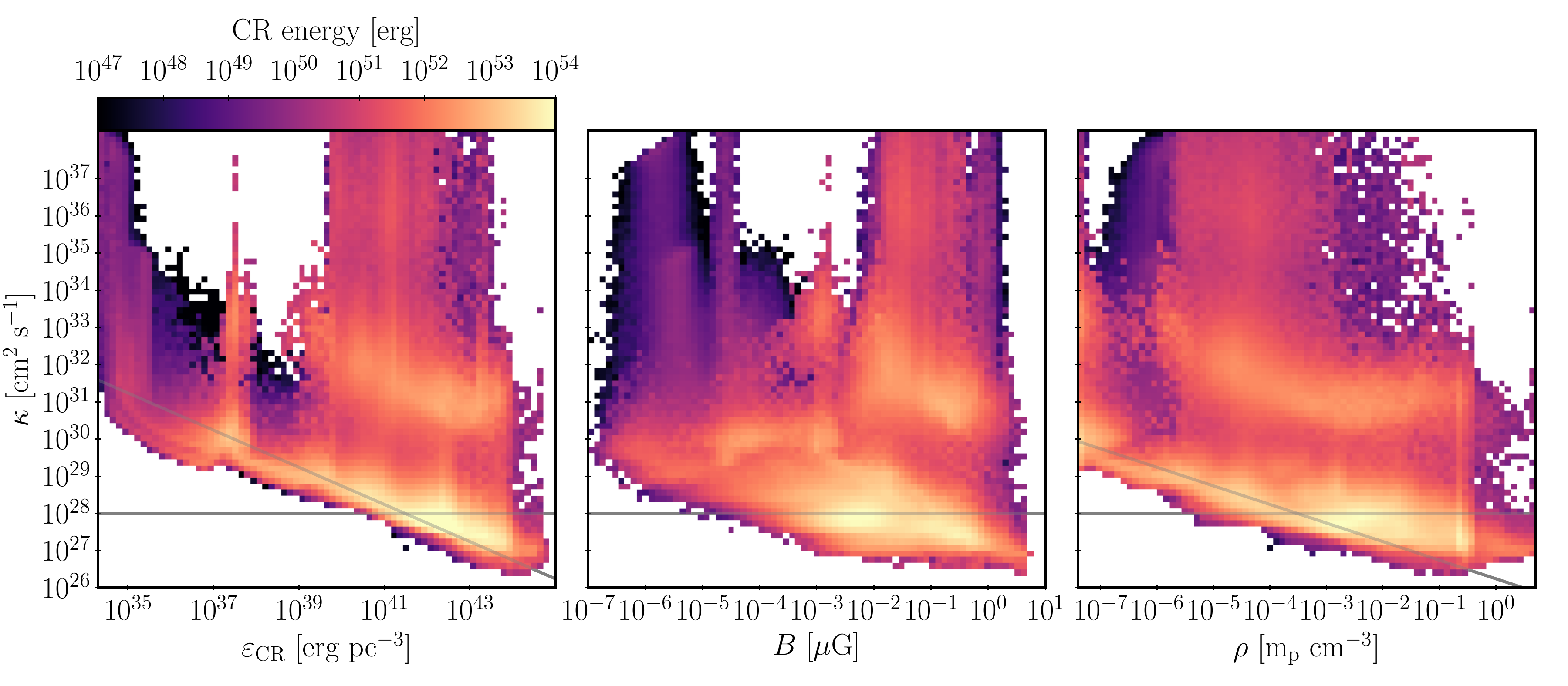}
	\caption{Scaling of the CR diffusion coefficient $\kappa$ (defined in Eq.~\ref{eq:total_diffusion_coefficient}) with the CR energy density (left panel), the magnetic field strength (middle panel) and the gas density (right panel) at $t = 4$~Gyr. We weight each computational cell of the simulation with its CR energy before binning into those histogram. The thick horizontal grey line marks the canonical value of the CR diffusion coefficient, $10^{28} \mathrm{cm}^{2} ~ \mathrm{s}^{-1}$. The thin oblique grey lines in the left and right panels display the correlations $\kappa \sim 10^{28} \mathrm{cm}^2 \, \mathrm{s}^{-1} (\varepsilon_\mathrm{cr} / 10^{42}\, \mathrm{erg} \, \mathrm{pc}^{-3})^{-0.5}$ and $\kappa \sim 10^{28} \mathrm{cm}^2 \, \mathrm{s}^{-1} (\rho / 3 \times 10^{-4}\, \mathrm{m}_\mathrm{p} \, \mathrm{cm}^{-3})^{-0.5}$.}
	\label{fig:diffusion_coefficient_correl}
\end{figure*}
equate growth and damping terms of Alfv\'en wave energy in Eqs.~\eqref{eq:forward_alfven_energy_equation} and \eqref{eq:backward_alfven_energy_equation} whilst assuming that CR energy is transported with $\varv_\mathrm{cr, steady}$. This leads to a steady state Alfv\'en wave energy of
\begin{equation}
    \varepsilon_\mathrm{a, steady} = \sqrt{\frac{\varv_\mathrm{a}}{\alpha} |\mathbfit{b}\bcdot \bnabla P_\mathrm{cr}|}
\end{equation}
which can be translated into an steady-state CR diffusion coefficient $\kappa_\mathrm{steady}$ using Eq.~\eqref{eq:def_kappa} to yield:
\begin{align}
    \kappa_\mathrm{steady} 
    &= \frac{4}{9 \pi^{3/4}}\left(\frac{e^2}{\gamma m c^4 P_\rmn{th}^{1/2}}\,\frac{P_\rmn{cr}}{L_\rmn{cr}}\right)^{-1/2}\\
    &\sim 9.2 \times 10^{27} \mathrm{cm}^{2}~\mathrm{s}^{-1}\nonumber\\
    &\times\left(\frac{\varepsilon_\mathrm{th}}{10^{42} \,\mathrm{erg \, pc}^{-3}} \right)^{1/4} \left(\frac{P_\mathrm{cr}}{L_\mathrm{cr}} \frac{1\, \mathrm{kpc}}{10^{42} \,\mathrm{erg \, pc}^{-3}} \right)^{-1/2},
    \label{eq:kappa_steady}
\end{align}
where $L_\mathrm{cr} = P_\mathrm{cr} / |\mathbfit{b}\bcdot \bnabla P_\mathrm{cr}|$ is the gradient length scale of the CR pressure. We compare $\kappa_\mathrm{steady}$ to the simulated non-steady state diffusion coefficient $\kappa$ in Fig.~\ref{fig:steady_kappa}. Each bin in this histogram is weighted by the sum of CR energy over all computational cells that fall inside it. Thus, a brighter colouring corresponds to a larger CR energy. We observe that most CRs scatter at a rate that is well described by a steady-state diffusion coefficient, $\kappa \sim \kappa_\mathrm{steady}$. We emphasise that both of these diffusion coefficients derive from the streaming and diffusion picture of CR transport where the streaming process provides an additional convective process that transports CRs along magnetic field lines. Both descriptions are not comparable to an effective diffusion coefficient $\kappa_\rmn{eff} = f_\rmn{cr} / (\mathbfit{b} \bcdot \bnabla \varepsilon_\rmn{cr})$ that describes the CR flux resulting from streaming and diffusion processes effectively in a diffusion-only approximation.

Both, Alfv\'en wave growth and damping are fast processes with typical time scales of $\sim 10$~kyr. Hence, we would naively expect the diffusion coefficient to also reach the steady state on these time scales. However, near the steady state the typical and effective time scale of Alfv\'en wave dynamics can be slower when damping and growth are nearly balanced. This leaves the probability of Alfv\'en waves with associated diffusion coefficients that only fluctuate around the steady state. 

In Fig.~\ref{fig:steady_kappa}, we find an additional population of CRs with $\kappa \sim 10^{3} \kappa_\mathrm{steady}$ and a broad distribution of diffusion coefficients that have  high $\kappa \gg \kappa_\mathrm{steady} \sim 10^{28} \mathrm{cm^2} \, \mathrm{s}^{-1}$. That particular sub-population of CRs and their associated Alfv\'en waves fail to reach a steady state. Inspecting Fig.~\ref{fig:kappa_gallery} we find that the broad distribution of high $\kappa$ values is caused by Alfv\'en wave dark regions and their vicinities where numerical diffusion causes a decrease of the surrounding Alfv\'en wave energy. This causes mixing of Alfv\'en-wave energies at the interfaces of dark regions and thus a broadened distribution of CR energy densities and the corresponding diffusion coefficients.

In Fig.~\ref{fig:diffusion_coefficient_correl} we correlate the simulated diffusion coefficient $\kappa$ with various quantities using two-dimensional histograms. We again weight each bin of the histogram with the CR energy contained to highlight the relevance of each bin for the CR dynamics. In the left-hand panel of Fig.~\ref{fig:diffusion_coefficient_correl} we compare the diffusion coefficient with the local CR energy density and also find distinct diffusion coefficients that correspond to the discussed steady-state population and Alfv\'en wave dark regions. The steady state population contains most of the CRs and has $\kappa \sim 10^{27}$--$10^{30} \mathrm{cm}^2 \, \mathrm{s}^{-1}$ for $\varepsilon_\mathrm{cr} \sim 10^{35}$--$10^{43} \mathrm{erg} ~ \mathrm{pc}^{-3}$. We find a weak correlation of $\kappa \sim 10^{28} \mathrm{cm}^2 \, \mathrm{s}^{-1} (\varepsilon_\mathrm{cr} / 10^{42}\, \mathrm{erg} \, \mathrm{pc}^{-3})^{-0.5}$ with a substantial scatter around the relation by approximately an order of magnitude. This scaling can be understood by assuming that the diffusion coefficient of these CRs is near its steady state value and thus can be described by Eq.~\eqref{eq:kappa_steady} while the typical length scale of CRs in the halo does not show large variations. In this case, $\kappa \propto \varepsilon_\mathrm{cr}^{-0.5}$ directly follows from Eq.~\eqref{eq:kappa_steady}. The second population of CRs is characterised by CR energy densities in the range $\varepsilon_\mathrm{cr} \sim 10^{40}$--$10^{43} \mathrm{erg} ~ \mathrm{pc}^{-3}$ but with $\kappa$ values that connect the steady-state diffusing CRs with $\kappa \rightarrow \infty$. CRs belonging to the second population reside in and around the Alfv\'en wave dark regions. Thus, steady-state diffusing CRs are not associated with the Alfv\'en wave dark regions and are consequently the volume filling population. 

Correlating $\kappa$ with the magnetic field strength $B$ in the middle panel of Fig.~\ref{fig:diffusion_coefficient_correl} reveals at most a very weak correlation between these quantities. Both, the volume filling CR population and the population associated with the Alfv\'en wave dark regions are visible. The magnetic field ranges four orders of magnitude for a given diffusion coefficient $\kappa$ of the volume filling CR population and thus, we cannot deduce a clear relation between the two quantities. For the volume filling population, where $\kappa \sim \kappa_\mathrm{steady}$, this follows from Eq.~\eqref{eq:kappa_steady} because $\kappa_\mathrm{steady}$ does not depend on $B$.

The right panel of Fig.~\ref{fig:diffusion_coefficient_correl} shows the relation of the diffusion coefficient with the local mass density $\rho$. The two populations of CRs are also visible and separated from each other. The diffusion coefficient of the volume filling CR population shows a strong scatter around a weak correlation with mass density, $\kappa \sim 10^{28} \mathrm{cm}^2 \, \mathrm{s}^{-1} (\rho / 3 \times 10^{-4}\, \mathrm{m}_\mathrm{p} \, \mathrm{cm}^{-3})^{-0.5}$. This scaling is empirical because $\kappa_\mathrm{steady}$ from Eq.~\eqref{eq:kappa_steady} does not directly depend on $\rho$. At mass densities that correspond to the halo-disc interface ($\rho \sim 10^{-3}$--$10^{-1} \mathrm{m}_\mathrm{p} ~ \mathrm{cm}^{-3}$) the typical diffusion coefficient is $\kappa \sim 3 \times 10^{27} \mathrm{cm}^{2} ~ \mathrm{s}^{-1}$.

We note that while the diffusion coefficient reaches a steady state most of the time, this is not the case for the CR flux as discussed in Section~\ref{sec:transport_velocity}. This difference is not a contradiction because the two quantities describe two separate physical processes that have different associated timescales. The diffusion coefficient is a measure for how fast CRs are scattered and its steady state is set by the balance of wave growth and wave damping. Consequently, this steady state can be reached on the timescales of the contributing growth and damping processes. The CR flux or the CR transport velocity are measures of how fast CR energy is transported and its steady state is mediated by the interaction with Alf\'en waves. This steady-state can only be reached on a much longer timescale that is not only characterised the CR scattering process but also by the hydrodynamic adjustments of the gradients in CR energy density, which appears to be the rate-limiting step (see Eq.~\ref{eq:vcr_steady}).

\section{Alfv\'en wave dark regions}
\label{sec:alfven-dark}

\begin{figure}
	\includegraphics[width=\linewidth]{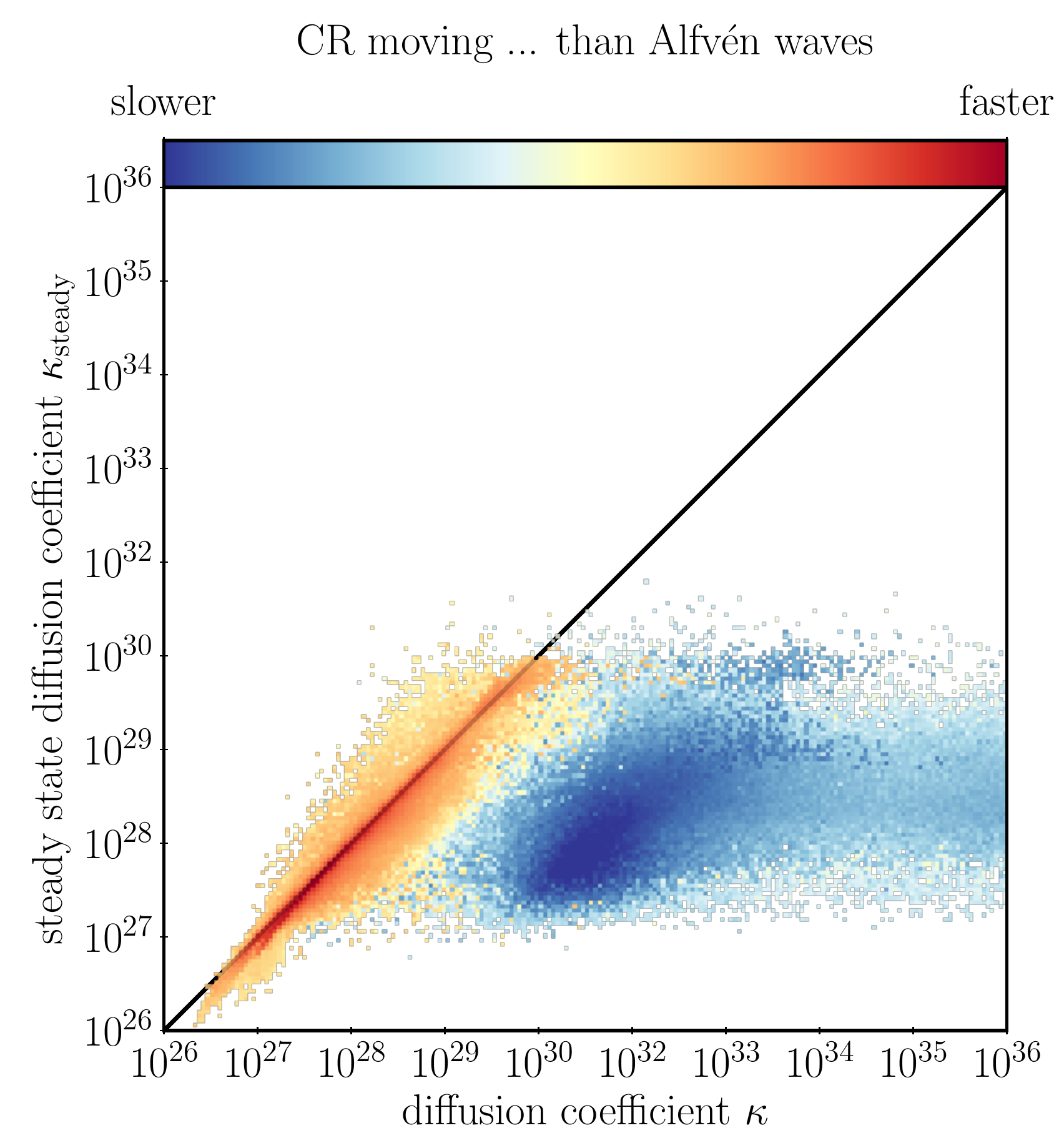}
	\caption{As Fig.~\ref{fig:steady_kappa} but here, the colour scheme distinguishes whether the CRs in a histogram bin preferentially move with a velocity $\varv_\mathrm{cr}$ that is slower (blue) or faster (red) than the Alfv\'en velocity. The shading of each colour (blue or red) is determined by the (logarithmic) amount of CR energy in that bin so that a more saturated colour indicates a larger average CR energy in that histogram bin.}
	\label{fig:steady_kappa_classification}
\end{figure}
In this section we discuss the origin of the Alfv\'en wave dark regions. They are characterised by a localised strong decrease of energy contained in gyroresonant Alfv\'en waves and a corresponding increase in the CR diffusion coefficient. We concluded in Section~\ref{sec:diffusion-coefficient} that the volume filling CR population is approximately transported with a steady-state diffusion coefficient while CRs in the Alfv\'en wave dark regions scatter considerably less frequent, which implies $\kappa\gg\kappa_\mathrm{steady}$. The simplest explanation for this situation is that the steady state, which is characterised by a balance of Alfv\'en-wave growth and damping, cannot be realised. Because the non-linear Landau damping damping rate scales linearly as $\Gamma_\mathrm{nlld} \propto \varepsilon_\mathrm{a,\pm}$, any wave growth could be balanced by this damping process because $\varepsilon_\mathrm{a,\pm}$ would adopt to match the corresponding wave growth rate. Thus, as long as CRs provide a positive growth rate through the gyroresonant instability, non-linear Landau damping should be able to balance wave growth and the Alfv\'en waves should settle into a steady state.

However, because a steady state cannot be reached for Alfv\'en wave dark regions, Alfv\'en wave growth is apparently prevented. This situation can be realised by CRs that move with $|\varv_\mathrm{cr}| < \varv_\mathrm{a}$. Those CRs drain energy from Alfv\'en waves through the gyroresonant interaction or, in other words, damp them. We can verify our conclusions by inspecting Fig.~\ref{fig:steady_kappa_classification} which shows the same data as Fig.~\ref{fig:steady_kappa} and thus correlates the steady state diffusion coefficient to the diffusion coefficient in our simulation. The difference between both figures is that Fig.~\ref{fig:steady_kappa_classification} colour-codes two categories: CRs in red-shaded colours move preferentially faster than Alfv\'en waves ($\varv_\mathrm{cr} > \varv_\mathrm{a}$) while CRs in blue-shaded colours move preferentially slower than Alfv\'en waves ($\varv_\mathrm{cr} < \varv_\mathrm{a}$). CRs with $\kappa \gg \kappa_\mathrm{steady}$ fall into the second category and are thus in a regime where they damp Alfv\'en waves. Within each category, we weight the distribution with the CR energy so that the colour saturation level represents the CR energy at a certain diffusion coefficient. CRs and associated Alfv\'en waves that show approximately steady-state values of the diffusion coefficient are transported super-alfv\'enically while CRs in the Alfv\'en wave dark regions are transported sub-alfv\'enically.
\begin{figure}
	\includegraphics[width=\linewidth]{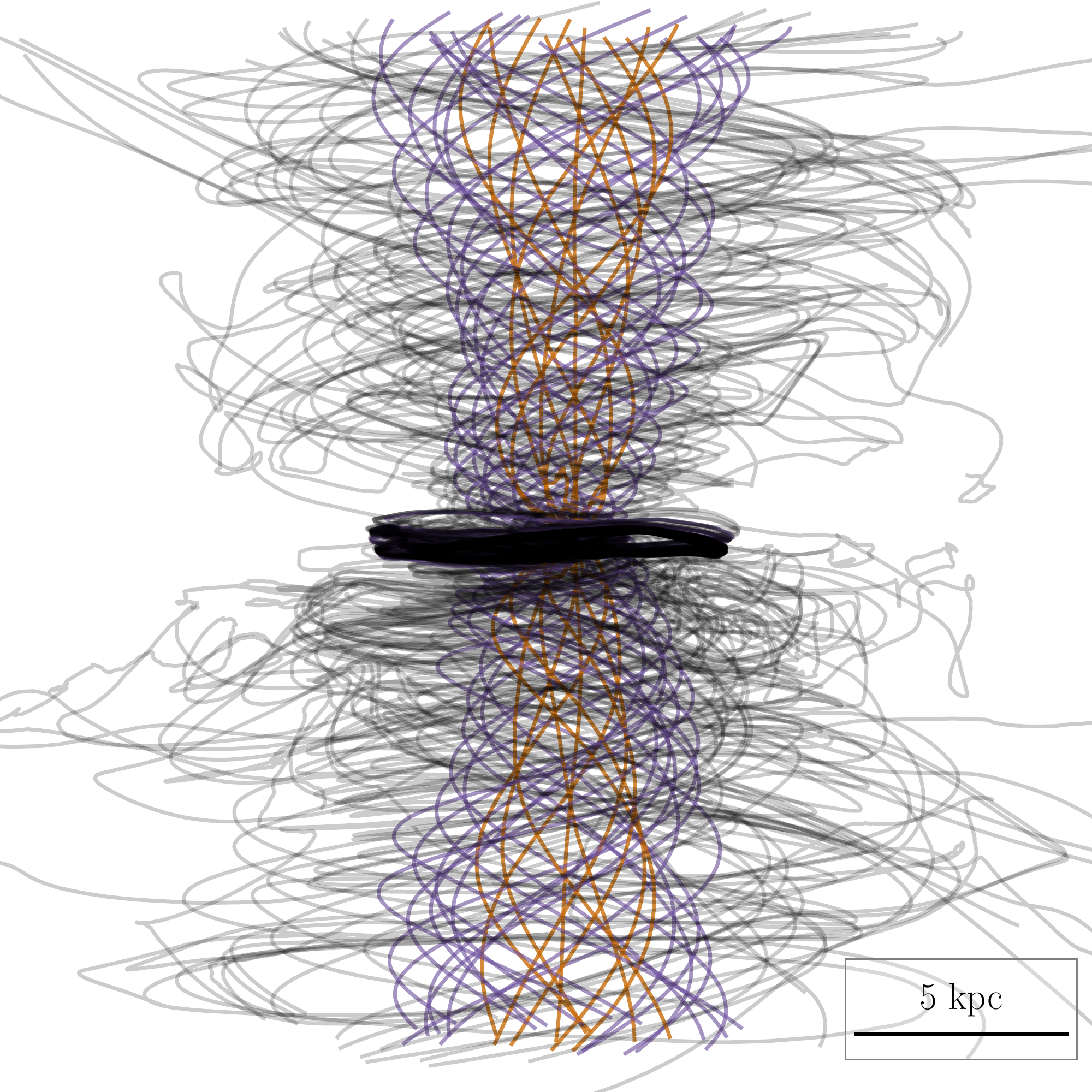}
	\caption{Individual magnetic field lines inside the inner CGM which anchor inside the galaxy and run through the inner galactic wind at $t = 4$~Gyr. The grey coloured field lines show the highly winded magnetic field lines in the outer wind, while the purple magnetic field lines separate the winded from the rather straight, orange-coloured field lines located in the inner wind.}
	\label{fig:magnetic_field_lines}
\end{figure}
\begin{figure}
	\includegraphics[width=\linewidth]{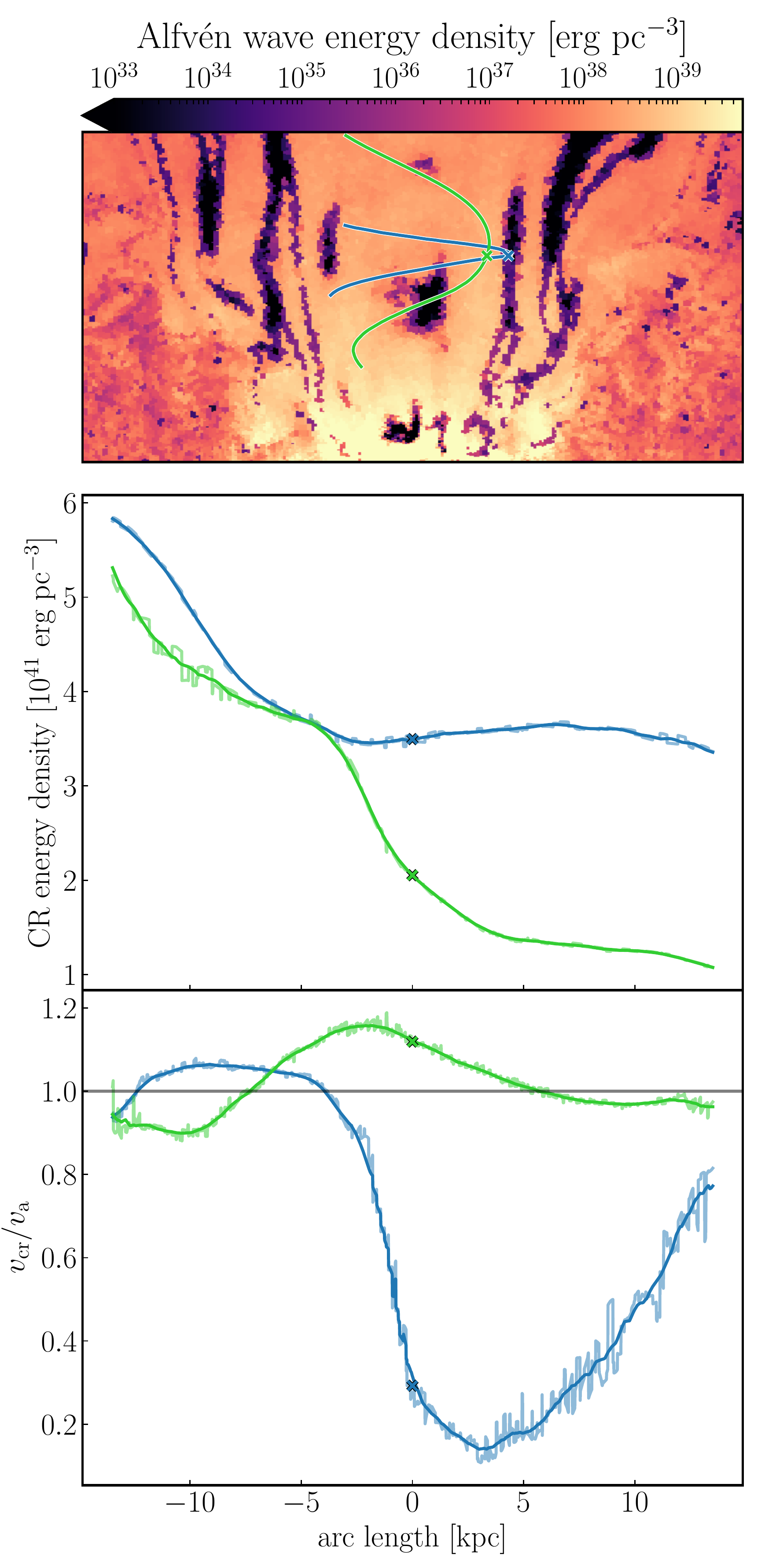}
	\caption{Top panel: same as the Alfv\'en wave energy density $\varepsilon_\mathrm{a} = \varepsilon_\mathrm{a,+} + \varepsilon_\mathrm{a,-}$ panel of Fig.~\ref{fig:gallery} but overlaid are the projections of two magnetic field lines onto the image layer. Both magnetic field lines intersect with the image plane at the location marked with a cross in the image. The line plots show the interpolated CR energy density $\varepsilon_\mathrm{cr}$ (middle panel) and the CR transport velocity in units of the Alfv\'en velocity $\varv_\mathrm{cr} / \varv_\mathrm{a}$ (bottom panel) along the two magnetic field lines. The crosses mark the position along the magnetic field line that corresponds to the intersection with the image plane in the top panel. The blue magnetic field line passes through a Alfv\'en wave dark region at the image layer while the green one does not. }
	\label{fig:ecr_along_fieldlines}
\end{figure}

Inspection of the bottom-left panel of Fig.~\ref{fig:gallery} or of Fig.~\ref{fig:kappa_gallery} reveals that the Alfv\'en wave dark regions are localised and isolated in the sense that there is no smooth transition between regions that are filled with or devoid of gyroresonant Alfv\'en waves along flowlines. These dark regions also appear to be clustered in the outer wind. This clustering also coincides with a transition of the magnetic field topology and hence a change of CR transport geometry from the inner to the outer wind. In Fig.~\ref{fig:magnetic_field_lines} we show magnetic field lines (generated by numerically integrating $\mathrm{d} \mathbfit{x} / \mathrm{d} \mathrm{s} = \mathbfit{B} / B$, where $s$ is the arc length along the magnetic field line) within the galactic outflow. The galaxy is visible as a dark accumulation of magnetic field lines in the centre of figure. Magnetic field lines are anchored in the disc but carried along with the galactic wind out of the ISM. This connects the galactic discs to the CGM by means of CR transport. The ratio of vertical-to-toroidal magnetic field component depends on the position inside the wind. Because the inner wind is faster, it is able to stretch the magnetic field lines in the vertical direction more easily. This leads to straighter appearance of magnetic field inside the inner wind. Contrarily, in the outer wind magnetic field lines are dominantly toroidal. As we move from the outer to the inner win, there is a smooth transition from the toroidal to the dominantly vertical magnetic field. We highlight magnetic field lines in each of the three different regions (inner wind, outer wind, and transition region) in Fig.~\ref{fig:magnetic_field_lines} with different colours. 

In Fig.~\ref{fig:ecr_along_fieldlines} we analyse the transport of CRs along two magnetic field lines. The blue magnetic field line was selected to pass through an Alfv\'en wave dark region (as indicated in the upper panel) while the green magnetic field line does not intersect an Alfv\'en wave dark region at the image layer. In the following, we call the points at which the field lines intersect the image panel \textit{points of interest}. We interpolate the CR energy density $\varepsilon_\mathrm{cr}$ and the ratio $\varv_\mathrm{cr} / \varv_\mathrm{a}$ along the two magnetic field lines and display these quantities in the lower panels of Fig.~\ref{fig:ecr_along_fieldlines}. We show the actually interpolated data (which contains noise due to the interpolation from the Voronoi grid onto the field lines) and a filtered version of the data which removes some of the small-scale noise. The horizontal coordinate is the arc length along the magnetic field line and negative arc lengths correspond to regions closer to the galactic disc.

We first focus on the \textit{green} magnetic field line. At its point of interest the CR energy density shows a monotonic, large gradient so that CR inertia is able to accelerate CRs beyond the Alfv\'enic point, $\varv_\mathrm{cr} \gtrsim \varv_\mathrm{a}$. This implies efficient driving of gyroresonant Alfv\'en waves, which leads to a sizeable Alfv\'en wave energy density as shown in the upper panel. The opposite happens along the \textit{blue} magnetic field line. Here, we identify an inflection point in $\varepsilon_\mathrm{cr}$ in the vicinity of the point of interest. Thus, although both magnetic field lines are embedded in the same global gradient of $\varepsilon_\mathrm{cr}$ (see central panel of Fig.~\ref{fig:gallery}), locally the gradient along $\mathbfit{B}$ is pointing in the opposite direction. Assuming that CRs are transported away from the galactic disc and towards the point of interest, the transport of CRs is rapidly decelerated at the point of interest due to CR inertia which results in $\varv_\mathrm{cr} < \varv_\mathrm{a}$. This in turn enables CRs to extract energy from gyroresonant Alfv\'en waves and leads to low levels of $\varepsilon_\mathrm{a}$ inside these regions.

\section{Discussion and caveats}
\label{sec:discussion}
In this work we show that the extension of CR hydrodynamics towards a more realistic description using our two-moment approach and a model for the interactions of CRs and self-generated small-scale Alfv\'en waves impacts the transport of CRs in a galactic wind. 

In our model we follow the CRs in the grey approximation and marginalise over the entire CR energy spectrum. This results in an effective description of CR transport with the overall CR energy budget being approximately dominated by GeV CRs \citep[assuming typical CR spectral indices,][]{2007Ensslin}. While this is certainly a valid first-order approximation, an improved description will follow the CR energy spectrum and thus be able to account for the energy-dependent transport of CRs. For example, \citet{2022Girichidis} consider the transport of the CR momentum spectrum and account for the variation of the CR diffusion coefficient with energy using the same collapsing halo setup as we do. They show that the inclusion of an energy dependent transport has implications for the wind launching and the distribution of CRs in the CGM. This is because CRs at different energies cool on different timescales and deviate in their transport speed and mode which leads to an overall change in the CR dynamics. Furthermore, because CR transport is tightly linked to the gas dynamics through momentum exchange, they find that the morphologies of the galactic wind and disc differ when simulated with spectrally resolved CRs. 

The CR transport velocity in our simulation is determined by CR interactions with gyroresonant Alfv\'en waves and consequently the processes that grow and damp these waves. Here, we only consider the gyroresonant interaction of CRs and non-linear Landau damping of Alfv\'en waves. While these are considered to be the most important processes, a plethora of other relevant processes that are associated with CR transport are not accounted for. Examples of those processes that may impact the dynamics of CRs in galactic winds are turbulent damping of Alfv\'en waves \citep{2004Farmer,2022Lazarian} but also additional scattering of CRs by magnetic fluctuations that are shifted into resonance through the turbulent cascade \citep{2011Yan}. Additionally, the scattering of CRs with small-scale ion-cyclotron waves through the intermediate scale instability \citep{2021Shalaby} might be important as it is able to faster exchange energy between CRs and plasma-scale electromagnetic waves in comparison to interaction with gyroresonant Alfv\'en waves. This can provide more CR scattering and consequently smaller CR diffusion coefficients.

Under realistic ISM conditions, SNe launch structured fountain-type outflows where cold ($\sim10$~K) and warm ($\sim10^4$~K) gas is entrained into a hot outflow. This is done in part by mixing hot halo gas with cold-warm gas to a thermally unstable phase that quickly cools and joins the warm phase so that there is a net growth of mass in this phase \citep{2018Gronke,2020Li,2020Sparre} as well as through magnetic tension of magnetic field lines that are anchored in the outflowing hot wind and draped around cold clouds \citep{2008Dursi}. Moreover, CRs that are accelerated at SN remnant shocks in the multiphase ISM and stream/diffuse away from the source regions into the CGM can also help to accelerate the wind and shape its properties \citep{2016Girichidis,2016Simpson,2018Farber,2018Girichidis}. 

The CR dynamics differs within cold-warm clumps in comparison to hot winds because non-linear Landau damping becomes subdominant as ion-neutral damping starts to prevail due to the availability of neutral atoms that collisionally damp gyroresonant Alfv\'en waves \citep{2021Armillotta}. This is clearly not only true for the galactic wind but also for the ISM itself. Thus, more detailed investigations of the dynamical impact of CR-Alfv\'en wave interactions on those small scales are welcome. While the cited ISM models with CR physics are simulating stratified boxes of localised ISM patches, only recently it has become possible to simulate and investigate the impact of individual SNe on galactic scales \citep{2017Hu, 2021Gutcke}, however limited to dwarf galaxies. An SN event in our simulations represents a population of SNe that inject CRs co-spatially and simultaneously into the ISM. For this reason and because we model the ISM through a modified equation of state that effectively describes the warm and hot phases with a single effective fluid, we are unable to locate the origin of the dynamics of CRs and accompanying Alfv\'en waves in spatially resolved, individual SNe. Because the diffusion coefficient is expected to strongly vary in space \citep{2021Semenov,2021Armillotta} and CRs impact the dynamics of the ISM, a thorough investigation of CR transport and their interplay with the surrounding environment in the multi-phase ISM is postponed to future work.

Our initial condition consists of a rotating gaseous halo. Due to cooling and gravity this halo collapses and forms a disc inside out owning to angular momentum conservation. Not all the gas directly collapses into the galactic disc but remains as a smooth and diffusive gaseous halo around the galaxy. The expelled galactic wind has to do volume work against the inertia of this halo gas. In a more realistic scenario where the galaxy is embedded in a cosmologically forming halo, the CGM is substantially more structured due to streams of accreting gas. A CR driven wind inevitably interacts with these gaseous streams, which alters gas flows inside the halo \citep{2020Buck}, the viral shock \citep{2021Ji}, and possibly observables of the CGM \citep{2021Butsky}. Applying our hydrodynamical scheme of non-equilibrium CR interactions with Alfv\'en waves in this setup would reveal whether the Alfv\'en wave dark regions we discovered percolate the entire galactic halo or whether they are only confined to the inner galactic wind.

\section{Conclusions}
\label{sec:conclusions}
Here, we study the formation and evolution of an isolated disc galaxy with halo mass $10^{11} \, \mathrm{M}_\odot$ that includes CR feedback through the injection of CR by SNe and CR transport based on the two moment description of \citet{2019Thomas}. This allows us to self-consistently follow crucial aspects of the transport of CRs along magnetic fields such as CR diffusion and streaming. Capturing the interactions between CRs and gyroresonant Alfv\'en waves in our hydrodynamical model enables us to self-consistently calculate the CR diffusion and scattering rates in a time and space resolved manner without the need to employ ad hoc choices for these coefficients. Our investigation focuses on the interaction of CRs with the galactic wind and the transport properties of CRs therein. Our results include:
 
\begin{itemize}
\item CR injected into the ISM start to build up a stratified distribution above and below the disc and the bulk of CRs gets accelerated by means of the CR inertia to super-alfv\'enic velocities at the halo-disc interface. If these CRs are efficiently coupled to the gas through their interaction with gyroresonant Alfv\'en waves, their momentum is transferred to the surrounding gas. This causes the CRs to be decelerated and the ambient gas to be accelerated by means of momentum-conservation, which eventually launches a galactic wind (see Figs.~\ref{fig:gallery} and \ref{fig:inner_profiles}).
\item The galactic outflow is bipolar, magnetised and is split into two characteristic regions. The inner wind is thermally dominated and underdense, which differs considerably from the outer wind, which is dominated by CR pressure. This outer wind interacts with the ambient CGM through shear motions and thus, turbulent mixing between the CGM and the wind leads to a more structured outflow morphology (see Fig.~\ref{fig:gallery}).
\item CR transport inside the galactic winds does not follow either CR streaming, CR diffusion or the mixed streaming~+~diffusion paradigm in the steady-state approximation. We find that CRs are preferentially transported in a non-steady state mode, which we characterise by a deviation from Fickian transport (see Figs.~\ref{fig:fickian_defect} and \ref{fig:velocity_classification}). This transport mode ultimately cannot be described by the popular one-moment CR transport paradigm and can only be captured using higher-order descriptions such as our two-moment method, which self-consistently solves for the effective transport speed of CRs that itself results from the dynamics and gyro-resonant Alfv\'en-wave scattering of CRs. 
\item The CR diffusion coefficient is a measure for the inverse CR scattering rate, which depends on the amplitude of Alfv\'en waves. In contrast to the CR transport speed (which cannot be described by steady state transport), we find that the CR diffusion coefficient in our simulation is approximated well for most of the CRs by the steady state value that is derived from balancing the growth and damping rates of Alfv\'en waves (see Fig.~\ref{fig:steady_kappa}). Correlating the CR diffusion coefficient with characteristic variables for CR transport reveals that $\kappa$ scales as $\varepsilon_\mathrm{cr}^{-0.5}$ and  $\rho^{-0.5}$ and shows no strong scaling with the local magnetic field strength (see Fig.~\ref{fig:diffusion_coefficient_correl}). 
\item Unexpectedly, we discover CRs with diffusion coefficients orders of magnitudes larger than the expected steady-state value. Those CRs populate isolated regions in the transition zone between the inner and outer wind, which are devoid of gyroresonant Alfv\'en waves and which we name Alfv\'en wave dark regions (see Figs.~\ref{fig:kappa_gallery} and \ref{fig:gallery}). This phenomenon is caused by CRs with effective transport speeds smaller than the local Alfv\'en speed which implies that CRs are extracting energy from Alfv\'en waves inside these regions (see Figs.~\ref{fig:steady_kappa_classification} and \ref{fig:ecr_along_fieldlines}).  
\end{itemize}
Our study of the plasma physics of CR transport in self-consistently driven galactic winds opens up the possibility for exploring how CR feedback impacts galaxy formation and evolution.

\section*{Acknowledgements}

TT and CP acknowledge support by the European Research Council under ERC-CoG grant CRAGSMAN-646955.

\section*{Data Availability}
The data underlying this article will be shared on reasonable request to the corresponding author.



\bibliographystyle{mnras}
\bibliography{main}



\appendix

\section{Exponential Integrator}

CR interactions with Alfv\'en waves are mediated through source terms in Eqs.~\eqref{eq:energy_equation} to \eqref{eq:backward_alfven_energy_equation}, which can be brought into the following algebraic form:
\begin{align}
    \frac{\mathrm{d} \mathbfit{U}}{\mathrm{d} t} = \mat{R}(\mathbfit{U}) \mathbfit{U},
    \label{eq:source_term_ode}
\end{align}
where the state vector is
\begin{align}
    \mathbfit{U} = \left(\varepsilon_\mathrm{cr}, f_\mathrm{cr}, \varepsilon_{a,+}, \varepsilon_{a,-} \right)^{T},
\end{align}
and the rate matrix is given by
\begin{align}
    \mat{R}\left(\mathbfit{U}\right) = \begin{pmatrix} \varv_\mathrm{a}^2 \chi \gamma_\mathrm{cr} T & - \varv_\mathrm{a} \chi D & 0 & 0 \\ c^2_\mathrm{red} \varv_\mathrm{a} \chi \gamma_\mathrm{cr} D & - c^2_\mathrm{red} \chi T & 0 & 0 \\ - \varv_\mathrm{a}^2 \chi \gamma_\mathrm{cr} \varepsilon_{a,+} & +\varv_\mathrm{a} \chi \varepsilon_{a,+} & - \alpha \varepsilon_{a,+} & 0 \\ -\varv_\mathrm{a}^2 \chi \gamma_\mathrm{cr} \varepsilon_{a,-} & -\varv_\mathrm{a} \chi \varepsilon_{a,-} & 0 & - \alpha \varepsilon_{a,-} \end{pmatrix}.
\end{align}
Here, we define the abbreviations
\begin{align}
    T &= \varepsilon_{a,+} + \varepsilon_{a,-}, \\
    D &= \varepsilon_{a,+} - \varepsilon_{a,-}, \\
    \chi &= \frac{3\pi}{8} \frac{e}{\gamma m c^3 B}.
\end{align}
In \citet{2021Thomas} we present a semi-implicit second order Runge-Kutta like scheme that numerically solves Eq.~\eqref{eq:source_term_ode} in combination with an adaptive timestepping algorithm to control the numerical error. The adaptive timestepping algorithm implements subcycling of the source term integration by dynamically increasing or the decreasing the integration timestep for a given subcycle until a given numerical error, which may be introduced through the numerical discretisation, is below an imposed limit. 

We found that the semi-implicit integrator of \citet{2021Thomas} causes the adaptive timestepping algorithm to request an unnecessary large number of subcycles to meet the criterion if the Alfv\'en wave energy densities are exponentially damped. This is the case when $\mat{R}(\mathbfit{U})$ is approximately constant and dominated by the entries in its lower-left corner. This behaviour is detrimental to the overall performance of the source term integration.
In order to increase the performance of the algorithm, we developed a drop-in replacement for the source-term integrator of \citet{2021Thomas}. The derivation is as follows: the analytic solution of Eq.~\eqref{eq:source_term_ode} is
\begin{align}
\label{eq:exp_solution}
    \mathbfit{U}(t) = \exp\left\{ \int_0^t \mathrm{d} t' \mat{R}\left[\mathbfit{U}(t')\right]  \right\} \mathbfit{U}(0),
\end{align}
which is a recursive solution because the argument of the matrix exponential itself depends on the solution $\mathbfit{U}(t)$. We can dissolve this recursive relation by applying a second-order Runge Kutta-like approximation to the integral in Eq.~\eqref{eq:exp_solution}. The result of this procedure is \begin{align}
    \mathbfit{U}^* &= \exp\left[ \Delta t \mat{R}(\mathbfit{U}^n) \right] \mathbfit{U}^n \label{eq:predictor_exp_integrator} \\
    \mathbfit{U}^{n+1} &= \exp{\left\{ \frac{\Delta t}{2} \left[\mat{R}(\mathbfit{U}^n) + \mat{R}(\mathbfit{U}^*)\right]  \right\}} \mathbfit{U}^n, \label{eq:final_exp_integrator}
\end{align}
where $\Delta t$ is the length of the timestep. This is a second-order numerical approximation to Eq.~\eqref{eq:source_term_ode}. We found that this integrator generally outperforms the original integrator of \citet{2021Thomas} in terms of the required computational resources to achieve a requested error tolerance in the adaptive timestepping algorithm. Hence, we use this new integrator for the simulation in this paper.

The advantage of the new integrator becomes apparent if we consider a situation that caused the problem of the old integrator in first place: if $\mat{R}(\mathbfit{U})$ is approximately constant or slowly varying then Eq.~\eqref{eq:source_term_ode} becomes a simple linear ordinary differential equation, which has an exponentially decaying or growing solution, depending on the sign of $\mat{R}(\mathbfit{U})$. In this situation the first stage in Eq.~\eqref{eq:predictor_exp_integrator} already provides a solution that is close to the analytical solution. If $\mat{R}(\mathbfit{U}) \sim \mathrm{const.}$, this first stage is already the exact analytical solution. If $\mat{R}(\mathbfit{U}) \neq \mathrm{const.}$ the second stage in Eq.~\eqref{eq:final_exp_integrator} is improving the numerical solution according to the predictor-corrector scheme. The new integrator returns a well-behaved numerical solution if $\mat{R}(\mathbfit{U})$ is slowly varying but encodes small physical timescales without triggering subcycling by the adaptive timestepping algorithm. This feature cannot be achieved with the old semi-implicit integrator.

\bsp	
\label{lastpage}
\end{document}